\documentclass[%
 reprint,
superscriptaddress,
 amsmath,amssymb,
 aps,
pra,
]{revtex4-1}

\usepackage[utf8]{inputenc}

\usepackage{float} 
\usepackage{graphicx}
\usepackage{dcolumn}
\usepackage{bm}
\usepackage{dsfont}

\usepackage{enumitem}
\newcommand{\da}{^{\dagger}}
\newcommand{\an}{\textbf{a}}
\newcommand{\ad}{\an\da}
\newcommand{\n}{n_{\mathrm{th}}}
\newcommand{\tr}{\mathrm{Tr}}
\newcommand{\sn}{\boldsymbol{\sigma}}
\newcommand{\sd}{\sn\da}
\newcommand{\sz}{\boldsymbol{\sigma_\mathrm{z}}}

\usepackage{xcolor}
\newcommand{\sam}[1]{{\color{black} #1}}

\begin{document}

\preprint{APS/123-QED}

\title{Probing a two-level system bath via the frequency shift of an off-resonantly-driven cavity}

\author{Thibault Capelle}
\affiliation{%
Laboratoire Kastler Brossel, Sorbonne Universit\'e,  CNRS, ENS-Universit\'e PSL, Coll\`ege de France, 75005 Paris, France}
\author{Emmanuel Flurin}
\affiliation{D\'epartement de Physique, ENS-Universit\'e PSL, CNRS, 24 rue Lhomond, F-75005 Paris, France}
\author{Edouard Ivanov}
\affiliation{%
Laboratoire Kastler Brossel, Sorbonne Universit\'e,  CNRS, ENS-Universit\'e PSL, Coll\`ege de France, 75005 Paris, France}
\author{Jose Palomo}
\author{Michael Rosticher}
\affiliation{D\'epartement de Physique, ENS-Universit\'e PSL, CNRS, 24 rue Lhomond, F-75005 Paris, France}
\author{Sheon Chua}
\author{Tristan Briant}
\author{Pierre-François Cohadon}
\author{Antoine Heidmann}
\author{Thibaut Jacqmin}
\author{Samuel Del\'eglise}
\affiliation{%
Laboratoire Kastler Brossel, Sorbonne Universit\'e,  CNRS, ENS-Universit\'e PSL, Coll\`ege de France, 75005 Paris, France}

\date{\today}

\begin{abstract}
\sam{}
Although the main loss channel of planar microwave superconducting resonators has been identified to be related to an external coupling to a two-level system (TLS) bath, the behavior of such \sam{resonators} in the presence of an off-resonant pump has yet to be fully understood. 
\sam{Alongside the well-known power-dependent damping, we observe a frequency shift with a conspicuous maximum for intermediate pump power that is attributed to a spectrally asymmetric saturation of the TLSs. We derive a semi-classical model that describes both of these effects quantitatively.}
The model is validated experimentally by performing a two-tone spectroscopy of several resonators fabricated on various substrates.
Together with the provided analytic formulas, the technique proposed here is a simple yet powerful tool to \sam{unambiguously identify the presence of a limiting TLS bath, and to} characterize various properties \sam{thereof}
, such as its average dephasing rate.

\end{abstract}

\maketitle

\section{Introduction}

Microwave superconducting resonators are an ubiquitous resource in various quantum devices, ranging from kinetic inductance bolometers\ \cite{Sauvageau1989,Audley1993} \sam{and} parametric amplifiers\ \cite{Roy2016,Pillet2015}, \sam{to} embedded circuits for cavity quantum electrodynamics\ \cite{Blais2004,Rau2004,Wallraff2004}. In addition, they play a central role as a microwave interface \sam{in} hybrid quantum systems such as solid-state spins\ \cite{Andre2006}, mechanical resonators\ \cite{Teufel2011,Toth2017}, and ferromagnetic magnons\ \cite{Tabuchi2014, Tabuchi2015}. Owing to the very low resistivity of superconducting materials, combined with advanced electromagnetic engineering \sam{to reduce} radiation losses, the dominant loss channel of such resonators is the dielectric loss due to the presence of a two-level system (TLS) bath in amorphous materials\ \cite{Wang2009}. A salient feature of this loss mechanism is its non-linear nature. Indeed, the damping of TLS-limited cavities was shown to depend on resonator occupancy, originating from thermal fluctuations\ \cite{Gao2008}, from resonant excitation\ \cite{ Lindstrom2009,Skacel2015}, or from non-degenerate resonant mode occupancy\ \cite{Sage2011,Katz2017}. \sam{The characterization of the microscopic properties of individual TLSs probed under stress\ \cite{Lisenfeld2015} or DC voltage bias\ \cite{sarabi2016} is an active field of research, and considerable} experimental efforts have been devoted to the characterization \cite{OConnel2008, Khalil2011, Vissers2012} and mitigation \cite{Paik2010, Wenner2011, Quintana2014, Bruno2015, Calusine2018} of TLS-related losses for superconducting resonators in the single photon regime. 
In many applications however, strong microwave tones are applied with significant frequency detuning from the resonance frequency in order to activate a parametric interaction between the resonator mode and \sam{various} degrees of freedom\sam{, such as a mechanical resonator \cite{Teufel2011}, an auxiliary microwave cavity \cite{Leghtas2015} or a superconducting qubit \cite{Lescanne2019}}. In this configuration, the off-resonant microwave tone responsible for the saturation of the TLS bath is spectrally distinct from the resonant mode subjected to the losses, \sam{simultaneously} giving rise to a shift of the resonance frequency and a modification of the population \sam{dependence} of the quality factor. 


In this letter, \sam{we propose a scheme} 
to probe the energy
relaxation and frequency shift properties of an off-resonantly-driven
microwave cavity due to the presence of a TLS bath. A semi-classical model that describes the modified susceptibility is derived and compared to experimental data obtained on several resonators fabricated on different substrates. Contrary to earlier works where a pump tone was injected at resonance with specifically engineered cavity modes\ \cite{Sage2011,Katz2017}, our technique is readily applicable to any kind of resonator \sam{, which makes it a versatile approach to identify a TLS loss mechanism, and quantify its contribution to the total resonator loss, a key capability for the design and optimization of high\nobreakdash-Q cavities}. Moreover, the ability to continuously scan the pump detuning allows us to determine the average dephasing rate of the TLS bath by relating it to a well-controlled experimental parameter.

The manuscript is organized as follows: in section \ref{sec:samples}, we describe the lumped-element resonators on which the measurements have been conducted. We then present in section \ref{sec:two_tone} the two-tone characterization scheme, and discuss a distinctive feature of this pump-probe technique: a pump-induced frequency shift of the resonator's frequency. In section \ref{sec:theory}, we propose a semi-classical model that accounts for the modified susceptibility of a microwave resonator due to the presence of a pump tone, and investigate numerically the validity of this model. Finally, in section \ref{sec:results}, we summarize the results obtained by fitting our model on various datasets obtained with different resonators fabricated on three different substrates. Together with the analytical formulas provided in section \ref{sec:theory}, the proposed two-tone technique constitutes a simple characterization method for the design of superconducting resonators, and a new tool for the investigation of amorphous materials.






\begin{figure}
\includegraphics[width=\linewidth]{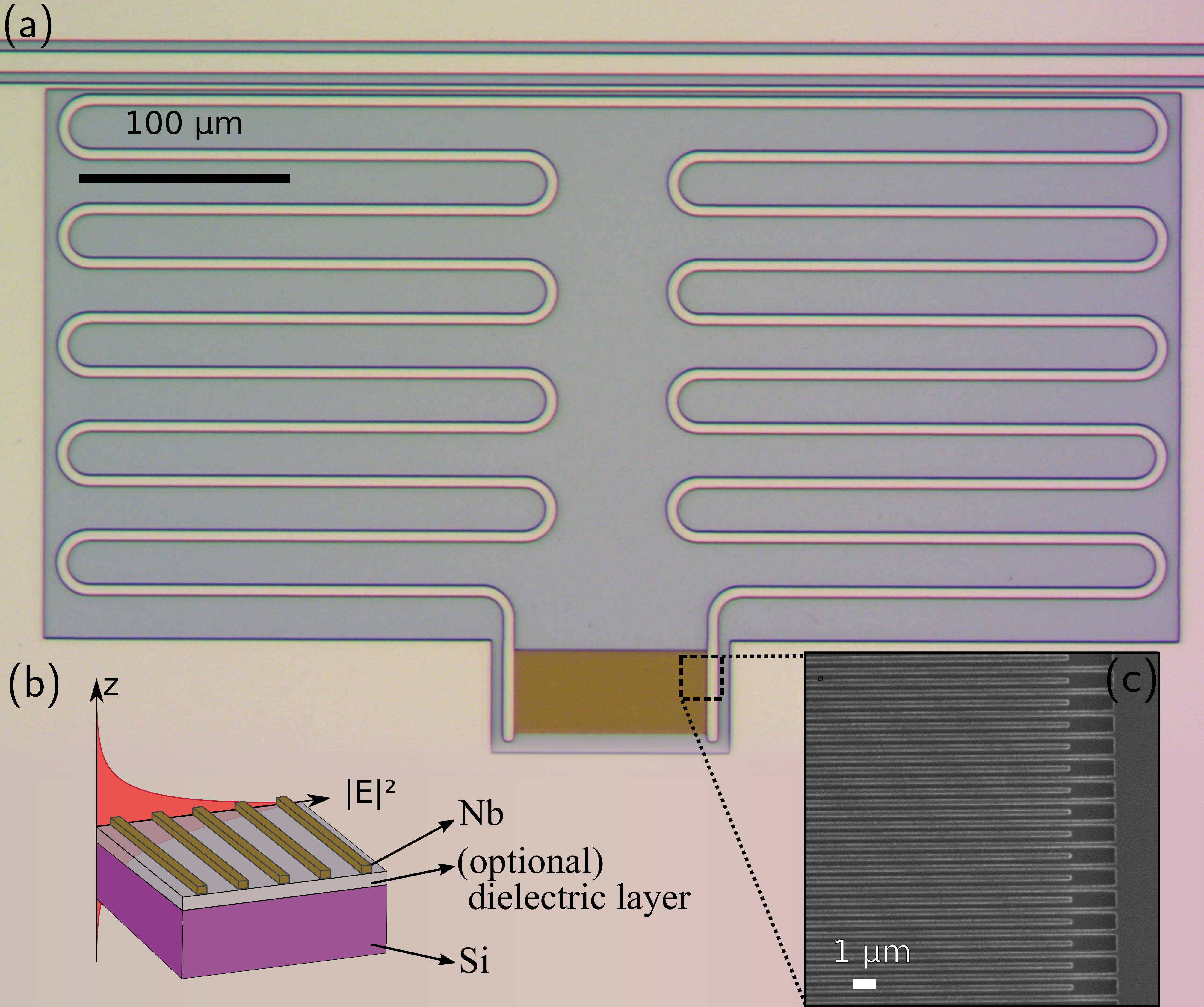}
\caption{\label{fig:sample} (a) Optical micrograph of the microwave cavity. (b) The dielectric environment of the capacitor and the profile of the electric field generated. Silicon is represented in purple, niobium is represented in grey, and the optional (SiO$_2$ or Si$_3$N$_4$) amorphous dielectric layer is represented in green. (c) view of the capacitor using electron microscopy.}
\end{figure}

\section{Sample design and fabrication}
\label{sec:samples}

Our samples are \sam{compact} resonators, made of a meander inductor in parallel to an interdigitated capacitor with a tooth period as small as 1~$\mu$m (see Fig. \ref{fig:sample}). These resonators have been initially optimized to maximize the coupling of the microwave mode to an external degree of freedom such as localized emitters \cite{Douce2015, Bertet2017} or the motion of a planar dielectric membrane placed in the evanescent field of the circuit via dielectric gradient forces \cite{Weig2009}. Due to the evanescent profile of the electric field around the interdigitated capacitor \cite{DenOtter2002}, the electric field is strongly confined in the direction transverse to the electrode plane. As a concomitant effect, the participation ratio of the surface interfaces and any deposited dielectric layers is thus enhanced in this geometry, which leads to an intentionally larger TLS signal in the form of resonator frequency shift and damping. Microwave radiation is coupled in and out of the resonator with a CPW feedline inductively coupled to the resonator.


To test the ability of our technique to discriminate between the properties of various dielectric materials, we fabricated resonators on 3 different substrates:  
\sam{a} 250 $\mu$m \sam{thick} substrate of \sam{float-zone} (FZ) grown (100) intrinsic silicon, with a resistivity of more than 10 000 $\Omega$.cm (substrate referred to as Si), 500 $\mu$m \sam{thick substrate} of FZ grown (100) P-doped type-n silicon, with a resistivity of more than 10 000 $\Omega$.cm, with 2 $\mu$m of SiO$_2$ from thermal oxidation (substrate referred to as Si/SiO$_2$), and finally, a 650 $\mu$m \sam{thick substrate} of Czochralski (CZ) grown (100) silicon, with P/\sam{b}oron doping and a resistivity of 1-30 $\Omega$ cm, with 200 nm of Si$_3$N$_4$ deposited through low-pressure chemical vapor deposition (substrate referred to as Si/Si$_3$N$_4$). 

The resonators are fabricated via e-beam lithography on a 100 nm thick niobium layer evaporated using a Plassys system under ultra-high vacuum ($\approx 5\times 10^{-10}$ mbar). For each susbtrate, 10 resonators have been  multiplexed on  the  same  coupling  waveguide  and their resonance frequency was staggered by incrementing the capacitor’s area by steps of 2.5 \%. The coupling-limited quality factor was designed to be $\approx 2\times 10^4$.

\section{Pump probe characterization}
\label{sec:two_tone}

\begin{figure*}
\includegraphics[width=0.8\linewidth]{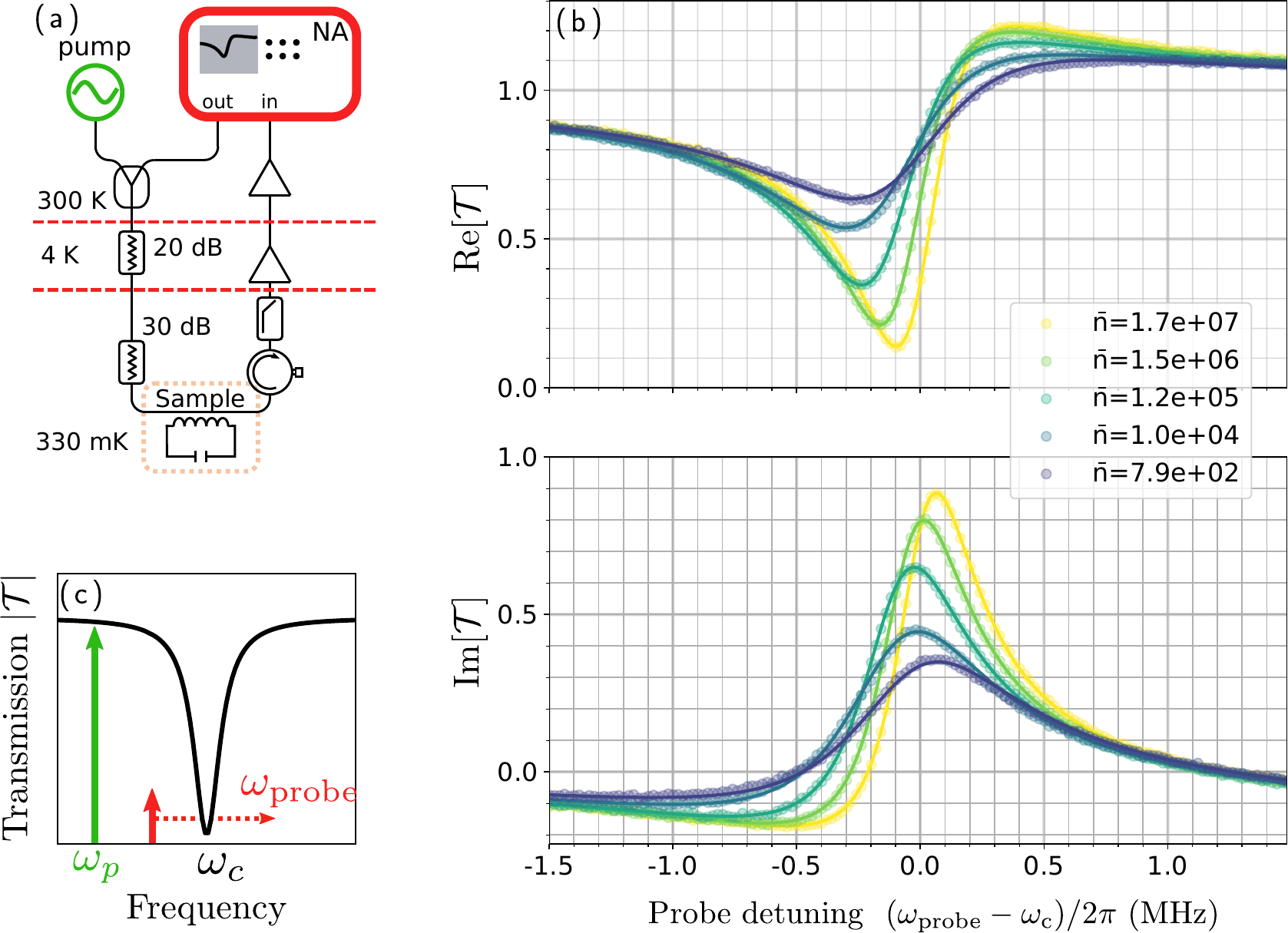}
\caption{\label{fig:setup}(a) Experimental setup: the output of a network analyzer (probe) is combined with that of a signal generator (pump) to drive the sample placed in a $^3$He cryostat. \sam{NA, network analyzer.} (b) Experimental protocol: a strong pump field is applied with a given detuning from a microwave cavity resonance, while a weak probe field is swept across the cavity resonance to measure its transmission spectrum $\mathcal{T}$. (c) Real (top) and imaginary part (bottom) of the probe transmission for a pump detuning of 4 MHz at various (pump) intracavity photon numbers. }
\end{figure*}

Each sample is placed in a $^3$He cryostat with a 330 mK base temperature, and probed by a two-tone excitation (see Fig. \ref{fig:setup}(a)): a strong pump with a fixed detuning in an interval spanning from several linewidths below to several linewidths above the cavity resonance, saturates the TLS bath and a weak probe is used to measure the resonance frequency $\omega_c$ and damping $\Gamma_\mathrm{tot}$ of the resonator (See Fig. \ref{fig:setup}(b)).  An example of the complex probe transmission $\mathcal{T}$ recorded for various pump power is represented in Fig. \ref{fig:setup}(c). Using a fit formula \cite{Geerlings2012, Khalil2012} that properly takes into account the effect of standing waves in the coupling waveguide, we can separate the contributions of the coupling waveguide $\Gamma_\mathrm{ext}$ and internal damping $\Gamma_\mathrm{int}$ in the total cavity linewidth $\Gamma_\mathrm{tot} = \Gamma_\mathrm{int} + \Gamma_\mathrm{ext}$.

Figure \ref{fig:results}(a) shows the frequency shift $\Delta \omega_c$ and internal damping $\Gamma_\mathrm{int}$ measured by the probe for various pump powers and detunings $\Delta$. Even with the large pump detuning, we observe a decrease of the resonator losses for increasing pump power. This effect can be attributed to the saturation of the TLS bath by the intracavity pump tone. 
To deconvolve the filtering of the detuned incoming pump tone by the cavity linewidth, the pump power is converted into intracavity photon number $\bar n$ via the formula $\bar n = 2 \Gamma_\mathrm{ext} |a_\mathrm{in}|^2/(\Gamma_\mathrm{tot}^2 + 4 \Delta^2)$, where $|a_\mathrm{in}|^2$ is the incoming photon flux in photon/s as determined by an independent calibration experiment (see Appendix C). Furthermore, to measure the correct intrinsic damping at low power, the probe beam was ensured to be sufficiently weak to prevent it from saturating the TLS bath.

A noticeable feature of this pump-probe experiment is the pump-dependent frequency shift observed in Fig. \ref{fig:results}(\sam{a}). This result is in contrast with single-tone experiments where the interaction with the TLS bath is only affecting the resonator damping. Moreover, the observed frequency shift has a non-monotonous behavior, with a maximum (respectively minimum) resonator frequency observed for a given pump power at positive (resp. negative) pump detuning $\Delta$. 

In the next section, we give a detailed theoretical description of the pump-probe experiment, and we propose an analytical model that can be used to link the previous experimental signatures with various properties of the bath, such as the average TLS dephasing rate.

\begin{figure*}
\label{fig:results}
\includegraphics[width=1.2\linewidth]{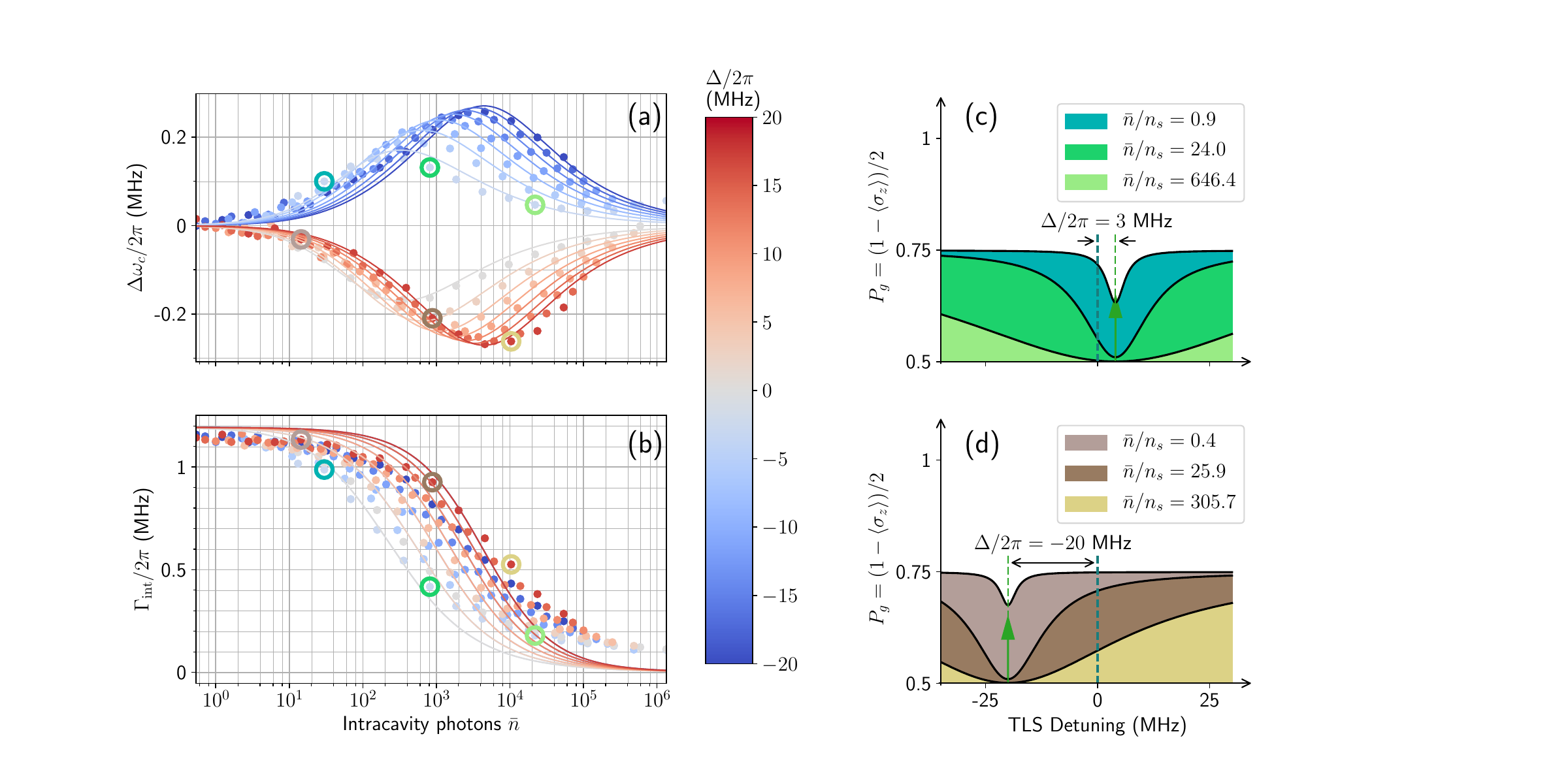}

\caption{\label{fig:results} (a) Frequency shift and (b) damping of the resonator, versus the number of intracavity photons. Each color corresponds to a different pump detuning ranging from $-20\,$MHz to $+20$\,MHz (color bar). Points correspond to experimental data, while solid lines correspond to fits by the model (Eq. \eqref{real_part_model} and \eqref{imaginary_part_model}). (c) and (d) are the ground state population distribution of the TLSs versus their frequency detuning $\omega_\mathrm{q} - \omega_\mathrm{c}$ calculated with Eq. \eqref{eq:population}. The various curves have been calculated for various intracavity pump photon numbers and the related points in (a) and (b) are highlighted with the corresponding color-circles. This measurement was performed on a  Si/SiO$_2$ sample, with $\omega_c/2\pi =$ 7.521 GHz (see circled point in Fig. \ref{fig:results_3d}).}
\end{figure*}

\section{Theoretical Analysis}
\label{sec:theory}

Here, we derive analytical formulas for the frequency shift and damping 
induced by an ideal bath of TLSs. These are characterized by a uniform 
frequency distribution of density
$P_0/2\pi$ (in Hz$^{\rm-1}$), a coupling $g$ to the resonator assumed to 
be identical for all TLSs in the distribution, and a damping (dephasing) 
rate $\Gamma_1$ ($\Gamma_2$). This is a simplifying assumption since the real TLS population has 
some statistical variation in these parameters (for instance, the 
coupling $g$ of individual TLSs to the resonator depends on the value 
and orientation of the electric field at the TLS location). However, by neglecting these effects, we can provide analytical formulas, which are numerically verified in
section \ref{sec:montecarlo} with a more  realistic TLS bath, and found to be in good agreement provided the fitted parameters are interpreted as averaged values over the TLS 
distribution.

\subsection{Analytical formula for a uniform TLS bath}

As the weak probe has a negligible effect on the TLS bath, the pump affects the population imbalance $\langle \sz(\omega_\mathrm{q}) \rangle$ according to the saturation law for a TLS at frequency $\omega_\mathrm{q}$
\begin{equation}
\label{eq:population}
\langle \sz (\omega_q) \rangle = \langle \sz \rangle_\mathrm{th} \left( 1 - \frac{\Gamma^2_2 \bar n/ n_\mathrm{s}}{(\omega_{\mathrm{q}} - \omega_{\mathrm{p}})^2 + \Gamma_{2}^2(1+\bar n/ n_\mathrm{s})} \right),
\end{equation}
where $\langle \sz \rangle_\mathrm{th} = -\tanh\left(\hbar \omega_\mathrm{q}/2 k_B T \right)\approx -0.52$ is the thermal imbalance resulting from the Fermi-Dirac distribution at the base temperature $T = 330$ mK of our $^3$He cryostat, $\Gamma_2$ ($\Gamma_1$) the TLS dephasing (energy relaxation) rate, $g$ the coupling rate, $\bar n$ the number of intracavity photons, $n_\mathrm{s}=\Gamma_1\Gamma_2/4g^2$ the number of photons required to saturate the TLS transition. In turn, the population imbalance $\langle \sz(\omega_{\mathrm{q}}) \rangle$ of a single TLS induces a shift of the complex cavity frequency \cite{Katz2017}
\begin{equation}
\label{TLS effect}
\delta\omega_c=\frac{g^2\langle\sz (\omega_{\mathrm{q}}) \rangle}{\omega_{\mathrm{q}} - \omega_\mathrm{c} + i\Gamma_2}.
\end{equation}
The frequency shift and damping, as measured by the probe beam, are related to the real and imaginary parts of $\delta\omega_c$. The total frequency shift and damping are obtained by integrating the contribution of individual TLS, assuming a flat spectral distribution of density $P_0$, uniform coupling rate $g$, and no interaction between individual TLSs, leading to (see Appendix B)
\begin{equation}
\label{real_part_model}
\Delta \omega_{\mathrm{c}} =
 - \frac{\Gamma_0}{2}\frac{\left(\Delta/\Gamma_2\right)  \left(\bar n/n_\mathrm{s}\right) }{\sqrt{1 + \bar n/n_\mathrm{s}} \left[\left(\Delta/\Gamma_2\right)^2 + \left( 1 + \sqrt{1 + \bar n/n_\mathrm{s}} \right)^2\right]} ,
\end{equation}
\begin{equation}
\label{imaginary_part_model}
\Gamma_\mathrm{int} =  \Gamma_0\left[1 - \frac{\bar n /n_\mathrm{s}}{\sqrt{1 + \bar n /n_\mathrm{s} }} \frac{1 + \sqrt{1 + \bar n /n_\mathrm{s}}}{\left(\Delta/\Gamma_2\right)^2 + \left(1 + \sqrt{1 + \bar n /n_\mathrm{s}} \right)^2} \right].
\end{equation}
In these formulas, in addition to the dephasing rate $\Gamma_2$, the TLS bath is described by two characteristic parameters: $\Gamma_0 = P_0 g^2 \left|\langle \sz\rangle_\mathrm{th}\right|$, the maximum damping produced by the TLS bath at the temperature $T$, that acts as a scaling factor on the curves $\Delta \omega_c (\bar n, \Delta)$ and $\Gamma_\mathrm{int} (\bar n, \Delta)$, and $n_\mathrm{s}$, that corresponds to a scaling of the curves with respect to the axis $\bar n$.

Eq. \eqref{real_part_model} and \eqref{imaginary_part_model} agree well with the experimental observations of the previous section. The solid lines in Fig. \ref{fig:results}(a) and (b) are a \sam{simultaneous} fit of the experimental points to Eq. \eqref{real_part_model} and \eqref{imaginary_part_model} with $\Gamma_0, \Gamma_2$, and $n_\mathrm{s}$ as free parameters. The small ($\sim 5 \%$) discrepancy between the measured damping and the fits at low pump power is attributed to the residual saturation of the TLS bath by the probe tone. 
We also observe a  residual loss at high pump power, that represents $\sim 10 \%$ of the low power value and that is attributed to a loss mechanism unrelated to the TLS bath. \sam{The smaller overall variation of the measured damping cannot be captured by the model since we perform a common fit on both damping and detuning points. We have nonetheless observed a decrease of this discrepancy as the probe power is weakened, consistent with the hypothesis of a residual saturation of the bath by the probe field. Since the detuning curve is not affected by these artifacts}
, we attribute a 90\,\% weighting to the detuning data in the \sam{global} fit
\sam{, and ignore these effects in our model for simplicity.} 
In the following sections, we give a qualitative explanation of the phenomena captured by our model and derive simple formulas in two limiting cases. 

\subsection{Small detuning limit}

When the pump detuning is small compared to the TLS dephasing rate ($\Delta \ll \Gamma_2$), the TLSs that are affected by the pump are the same as in a single-tone experiment. In particular,
since an equal number of TLSs are excited by the pump on either side of the cavity, the frequency shift vanishes in this regime:
\begin{align}
    \Delta \omega_c &= 0, \\
     \Gamma_\mathrm{int} &=  \frac{\Gamma_0}{ \sqrt{1 + \bar n/n_\mathrm{s}}}.
    \label{eq:damping_single_tone}
\end{align}
Eq. \eqref{eq:damping_single_tone} is the well known power-dependent absorption of the TLS bath derived in the context of single-tone  experiments \cite{Phillips1987}.

\subsection{Large detuning limit}
In the large detuning limit ($\Delta \gg \Gamma_2$), the effect of the pump field on the TLS distribution is more subtle: as the intracavity field resonates at a frequency $\omega_p$ that is significantly different from $\omega_c$, a depletion in population imbalance $\langle \sz(\omega_{\mathrm{q}}) \rangle$ occurs for TLSs that have a frequency $\omega_q$ close to $\omega_p$. More quantitatively, the width of this Lorentzian dip is given by the generalized Rabi frequency $\Gamma_2 \sqrt{1 + \bar n/n_\mathrm{s}}$. Consequently, the frequency pull exerted by TLSs that are above the cavity frequency will not be perfectly compensated by those that are below, resulting in a net shift of the cavity resonance. At even larger pump power, the width of the dip in population imbalance exceeds the pump detuning, such that the asymmetry decreases.  Qualitatively, the maximum cavity frequency shift occurs when the pump creates a depletion of population imbalance $\langle \sz(\omega_{\mathrm{q}}) \rangle$ of spectral width $\Delta$. In the current limit, this occurs when $\bar n \approx n_s \left(\Delta/\Gamma_2\right)^2$. This effect is illustrated in Fig. \ref{fig:results}(c) and (d), where the ground state population has been calculated using \sam{Eq.} \eqref{eq:population} for two different pump detunings, and various pump powers spreading below and above this value.

Further, we can note that \sam{Eqs.} \eqref{real_part_model} and \eqref{imaginary_part_model} can be approximated in the large detuning limit by:
\begin{align}
    \Delta \omega_c &= - \frac{\Gamma_0}{2} \frac{\delta}{\delta^2 + 1} \\
    \Gamma_\mathrm{int} &= \Gamma_0 \frac{\delta^2}{1 + \delta^2}
\end{align}
with the dimensionless parameter $\delta =\sqrt{n_\mathrm{s}/\bar n} \Delta / \Gamma_2$. Hence, in the regime $\Delta \gg \Gamma_2$, the curves $\Delta \omega_c(\bar n, \Delta)$ are invariant under the transformation $(\bar n, \Delta) \rightarrow (\sqrt{\alpha} \bar n, \Delta/\alpha)$, where $\alpha$ is an arbitrary positive number. In particular, the maximum frequency shift and damping are independent of the pump detuning provided it greatly exceeds the dephasing rate $\Gamma_2$ of the bath and their ratio is a non-adjustable prediction of the model:
\begin{equation}
    \frac{\Delta\omega_c(\delta = 1)}{\Gamma_\mathrm{int}(\bar n = 0)} = 1/4.
\end{equation}
We find experimentally $\mathrm{max}(\Delta\omega_c)/\mathrm{max}(\Gamma_\mathrm{int}) \approx 0.23$ with the data presented in Fig. \ref{fig:results}(a) and (b).

By capturing the transition between these two distinct regimes, the fit with the full Eqs. \eqref{real_part_model} and \eqref{imaginary_part_model} performs a direct comparison between the TLS bath dephasing rate $\Gamma_2$, and the known pump detunings $\Delta$. The two-tone experiments therefore give direct experimental access to $\Gamma_2$, a parameter that is elusive to the single-tone probing of a TLS bath.

\subsection{Effect of the non-uniform TLS distribution}
\label{sec:montecarlo}

\begin{figure*}
\label{fig:monte_carlo}
\includegraphics[width=\linewidth]{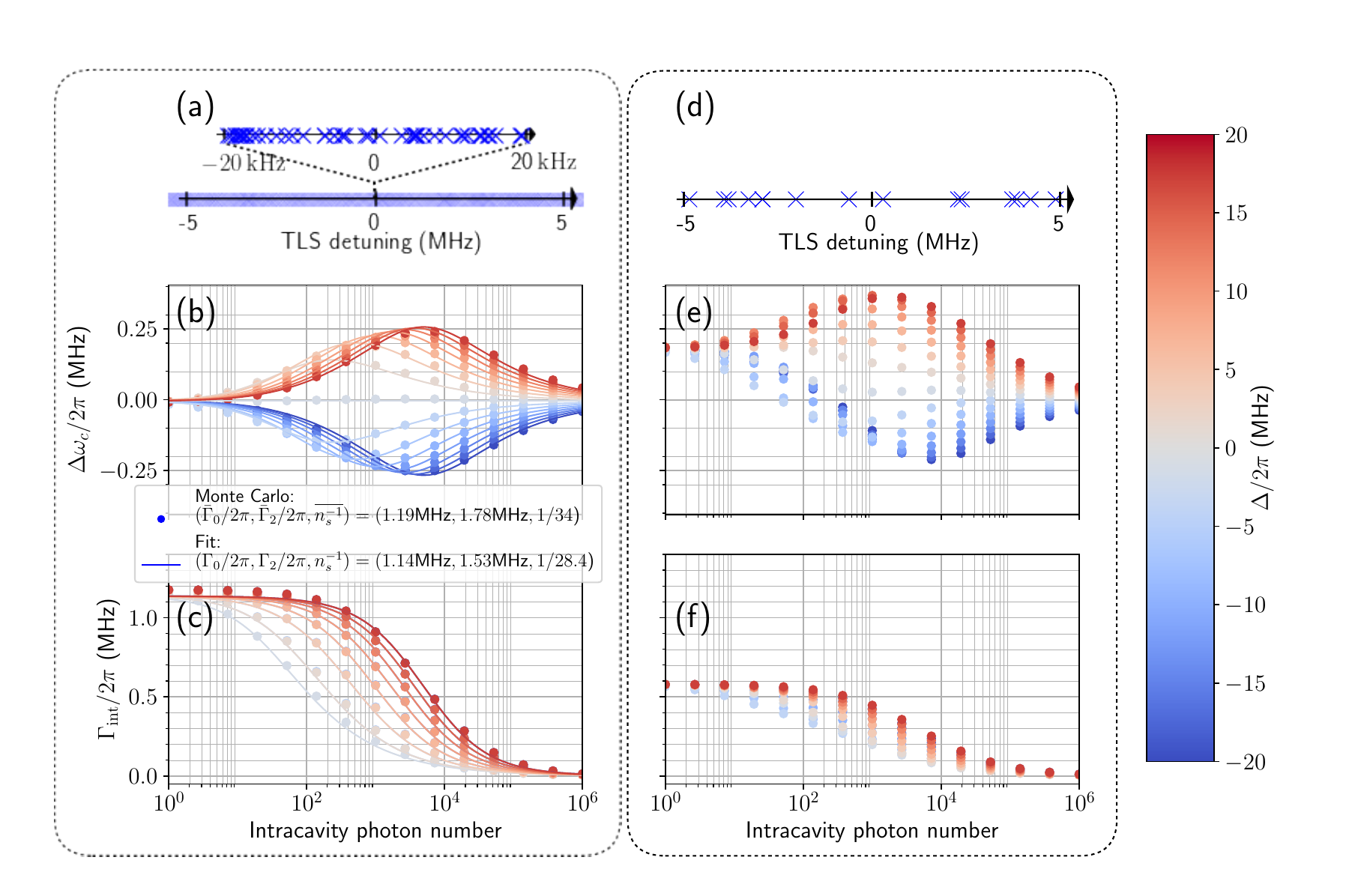}
\caption{\label{fig:monte_carlo}  (Left) Numerical validation of the model:  the points in (b) and (c) are the frequency shifts and dampings calculated with a Monte Carlo simulation for various pump power and detuning. The TLS frequencies are sampled from a flat distribution of density $P_0/2\pi=1\,$kHz$^{-1}$. A 10 MHz fragment of the sampled TLS frequencies is visible as a collection of blue crosses in (a), as well as a zoom on a 40 kHz-wide region. The average values $(\bar \Gamma_0, \bar \Gamma_2, \overline{n_s^{-1}})$ of the sampled population (see legend) match the values obtained with the experimental fit of Figure \ref{fig:results}. The full lines are fits with Eqs. \eqref{real_part_model} and \eqref{imaginary_part_model} (fitted parameters are \sam{also} indicated in the legend). (Right) Random sampling effects at low TLS density: the points in (e) and (f) are the typical Monte Carlo results for the same parameters as in (b) and (c), except for a lower spectral density $P_0/2\pi = 1$ MHz$^{-1}$. In this particular realization, the reduced damping and positive frequency shift at low pump power results from a deficit of TLSs on the low frequency side of the cavity (see sampled TLS frequencies over a 10 MHz-fragment in (d)).}
\end{figure*}

\subsubsection{Monte Carlo simulations}
The model derived in the previous section is based on the assumption that the bath is composed of a large number of TLSs with identical properties. In this section, we study numerically how a non-uniform distribution of TLS parameters affects the previous findings. We perform a Monte Carlo simulation where we randomly pick an ensemble of $N$ TLSs characterized by the parameters $\{\omega_{q,i}, g_{i}, \Gamma_{1,i}, \Gamma_{2,i}\}_{i \in [1..N]}$. The individual TLS frequencies $\omega_{q,i}$ are drawn from a uniform distribution of density $P_0$. The couplings ${g_{i}}$ are chosen randomly in a uniform distribution on the intervals $[0, g_\mathrm{max}]$: this distribution would be rigorously justified for a bath of TLSs with random orientations in a uniform electric field. The energy damping and dephasing rates  $\Gamma_{1,i}$ and $\Gamma_{2,i}$ are drawn from a log-normal distribution, where the standard-deviation of the variable's logarithm is fixed to 1/2.

The TLS spectral density $P_0$ is not constrained by our model, we thus choose a starting value $P_0/2\pi = 1 $ kHz$^{-1}$ large enough to ensure that the numerical results are insensitive to the sampling noise associated with the random realization of TLSs. In practice, we choose $10^6$ TLSs in a 1 GHz interval around the cavity frequency. Moreover we choose the mean values of the probability distributions such that $\bar \Gamma_0 = \left| \langle \sz\rangle_\mathrm{th} \right| P_0 \langle {g_{i}^2} \rangle$, $\bar \Gamma_2 = \langle \Gamma_{2,i} \rangle$, and $\overline{n_s^{-1}}= \langle \frac{4 g_i^2}{\Gamma_{1,i} \Gamma_{2,i}} \rangle$ match the values fitted with our model on the experimental data of Fig. \ref{fig:results}.

The expected frequency shift and damping are then computed on a regular grid of pump detuning and intracavity power by summing the contribution of individual TLSs using Eq. \eqref{TLS effect}. The resulting graph is represented in Fig. \ref{fig:monte_carlo}, together with a fit using the analytic Eqs. \eqref{real_part_model} and \eqref{imaginary_part_model}. We observe a very good agreement between the fits and the values calculated with the simulations; the fitted values of $\Gamma_0, \Gamma_2$ and $n_s^{-1}$ match to within 15 \% the average values $\bar \Gamma_0, \bar \Gamma_2$ and $\overline{n_s^{-1}}$ of the distributions sampled in the Monte Carlo simulation. This indicates that although the Eqs. \eqref{real_part_model} and \eqref{imaginary_part_model} were rigorously derived with the assumption of a unique value of $g^2$, $\Gamma_1$ and $\Gamma_2$, they well describe the effect of a non-uniform TLS bath, provided the fitted values are interpreted as average values over the TLS distribution.


\subsubsection{Effect of random sampling}

Our model is insensitive to the spectral density of TLSs: the scaling $(P_0, g, \Gamma_1)\rightarrow (\alpha P_0, \alpha^{-1/2} g, \alpha \Gamma_1)$ with $\alpha$ an arbitrary positive number, leaves the parameters $\Gamma_0, \Gamma_2, n_s$ unchanged. However, the smaller the density $P_0$, the smaller the number of resonant TLSs will contribute to the complex frequency pull. When only a handful of TLSs contribute to the effect, we observe the signatures of the random sampling in the Monte Carlo simulation. Fig. \ref{fig:monte_carlo} (e), (f) shows the typical shape of $\Delta \omega_c (\bar n, \Delta)$ and $\Gamma_\mathrm{int} (\bar n, \Delta)$ for a TLS density as low as $1$ MHz$^{-1}$. We observe an asymmetry in the frequency shift of the resonator: at low pump power, the frequency shift is mainly governed by the few TLSs that are located within a frequency difference $\Gamma_2$ from the resonator. An excess on one side of the cavity leads to a constant shift of the cavity frequency. On the other hand, in the large pump power limit, the cavity recovers its unshifted frequency since all the TLSs in a large frequency span around the cavity are saturated. Depending on the particular frequencies of the TLSs close to the cavity resonance, the low-power shift can be either towards low or high frequency. These effects have not been observed in our experiments, and we thus conclude that the TLS density $P_0/2\pi \gg 1$~MHz$^{-1}$. This result is consistent with other work from the literature \cite{Burnett2016}, that found a typical surface density for resonant TLS of $\sim$ 1\,$\mu$m$^{-2}$ \sam{(resonant TLSs are those with a detuning $|\omega_{\mathrm{q},i}-\omega_c|\lesssim \Gamma_2$, that contribute significantly to the low-power damping effect)}. With this estimate, we can infer that approximately 5000 TLSs contribute to the resonator shifts\sam{, or equivalently, a frequency density $P_0/2\pi$ in the kHz range.}

\section{Measurement results}
\label{sec:results}

\begin{figure*}
\label{fig:results_3d}
\includegraphics[width=\linewidth]{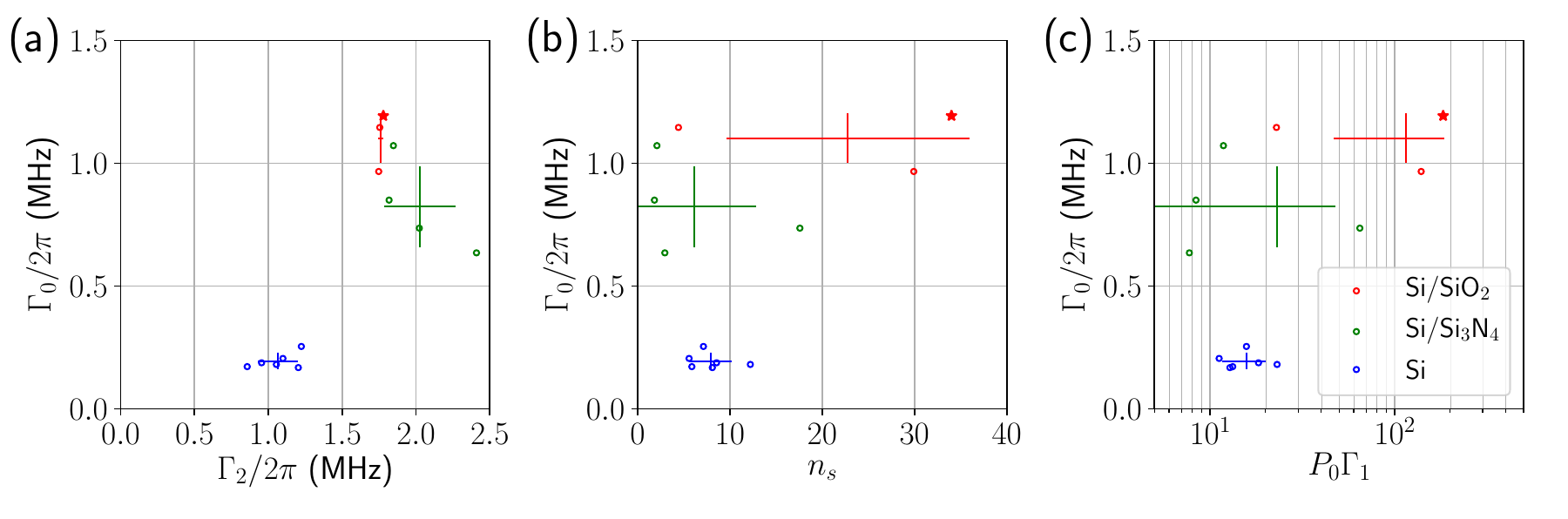}
\caption{\label{fig:results_3d}
\sam{Fit results: each point in the scatter plots represents the result of a fit similar to that in Fig. \ref{fig:results}(a), (b). Resonators fabricated on a Si/SiO$_2$, Si/Si$_3$N$_4$, and Si substrates are represented by red, green, and blue points respectively. The coordinates ($\Gamma_0, \Gamma_2$) and ($\Gamma_0, n_s$) extracted on each sample are represented in (a) and (b) respectively. The same data-points are represented in the ($P_0 \Gamma_1, \Gamma_0$) plane in (c), see text for details. The mean and standard deviation of each ensemble is represented as thick crosses in each of the plots. The starred data point is the result of the fit presented in Fig. \ref{fig:results}(a) and (b).}}
\end{figure*}

To evaluate the dispersion in the parameters estimated by our technique, we have repeated the fit presented in Fig. \ref{fig:results} on various resonators. Out of the 10 resonators fabricated on each of the Si/SiO$_2$, Si/Si$_3$N$_4$, and Si substrates, we have observed 3, 4, and 6 resonances respectively. We attribute the missing resonances to the presence of \sam{short circuits} in the interdigitated capacitors.

\sam{The 3 parameters $(\Gamma_0, \Gamma_2, n_s)$ extracted from the fit of each operative resonator are represented as a point in the $(\Gamma_2,\Gamma_0)$ and $(n_\mathrm{s},\Gamma_0)$ planes in Fig. \ref{fig:results_3d}(a) and (b) respectively.} The resonators fabricated on Si/SiO$_2$, Si/Si$_3$N$_4$ and Si substrates are represented in red, green, and blue points respectively. For a given substrate, the standard deviation of each parameter is represented as an error bar on the corresponding \sam{plot}. \sam{T}able \ref{tab:results} also summarizes the values extracted and the corresponding standard deviations for the different substrate types. The large dispersion of the parameter $n_s$, in particular for the Si/SiO$_2$ and Si/Si$_3$N$_4$ substrates, is likely due to systematic errors in the calibration of $\bar n$. Indeed, the determination of $\Gamma_\mathrm{ext}$ is difficult for these largely undercoupled resonators (for some resonators, we have found a contribution of the coupling to the waveguide as low as $\sim 3$ \% of the total damping). However, since this effect only corresponds to a global shift of the curves $\Delta \omega_c(\bar n, \Delta)$, $ \Gamma_\mathrm{int}(\bar n, \Delta)$ towards lower occupancies $\bar n$, the determination of  $\Gamma_0$ and $\Gamma_2$ is not affected by this inaccurate calibration. The losses $\Gamma_0$ are typically higher by a factor 5 with the Si/Si$_3$N$_4$ and Si/SiO$_2$ substrates as compared to Si. This indicates that most of the TLSs are indeed located in the amorphous layers SiO$_2$ and Si$_3$N$_4$. Moreover, we find a dephasing rate $\Gamma_2/2\pi$ in the MHz range for the 3 kinds of substrate, with a consistent variation from $1.07 \pm 0.13$ MHz for Si substrates, $1.76 \pm 0.013$ MHz for the SiO$_2$ substrates, and $ 2.02 \pm 0.24$ MHz for the Si$_3$N$_4$ layers. This points towards different TLS microscopic nature in the various materials. \sam{Even though the TLS spectral density $P_0$ cannot be resolved by our technique, a \emph{normalized} TLS spectral density $P_0 \Gamma_1 = 4 \Gamma_0 n_\mathrm{s}/\Gamma_2 |\langle \sz \rangle_\mathrm{th}|$ can be calculated from the fitted parameters $(\Gamma_0, \Gamma_2, n_\mathrm{s})$. Fig. \ref{fig:results_3d}(c) represents the same experimental points as in Fig. \ref{fig:results_3d}(a) and (b) in the ($P_0 \Gamma_1$, $\Gamma_0$) plane. For the three substrate types the normalized TLS spectral density is comprised between $\sim$10 and $\sim 200$ TLSs per intrinsic linewidth. As for the parameter $n_\mathrm{s}$, the relatively large scatter observed for the Si/Si$_3$N$_4$ and Si/SiO$_2$ substrates results from the poor determination of $\Gamma_\mathrm{ext}$ for these resonators.
The lowest values are found on the Si and Si/Si$_3$N$_4$ substrates and the largest values on the Si/SiO$_2$ substrates. The low spectral density of TLSs on the Si/Si$_3$N$_4$ (comparable to the bare Si substrate) is likely due to the small thickness of the amorphous Si$_3$N$_4$ layer on these samples: the TLSs contributing to the damping are relatively rare, but strongly coupled to the resonator due to their proximity to the interdigitated capacitor.}

The low-power internal quality factor $Q(\mathrm{330\,mK}) = \omega_c/\Gamma_0$, and the quality factor extrapolated at zero temperature and zero pump power $Q(\mathrm{0\,K})=Q(\mathrm{330\,mK}) \left| \langle \sz \rangle _\mathrm{th} \right|$ is also presented in separate columns. The relatively low value of $\sim 10^4$ even for the samples fabricated on Si is likely due to the unusually small pitch of the interdigitated capacitors studied here \cite{Geerlings2012}, along with the absence of a chemical surface treatment prior to metal deposition \cite{Bruno2015}.

\begin{table*}[tbh]
\begin{center}
\begin{tabular}{|c |c |c |c| c| c| c|}
\hline
Type of wafer&$\Gamma_0/2\pi$ (MHz)& $\Gamma_2/2\pi$ (MHz)& $n_\mathrm{s}$ (ph) & $P_0 \Gamma_1$ & Q(330 mK) & Q(0 K)\\
\hline
Si/SiO$_2$& $1.1 \pm 0.098$& $1.76 \pm 0.013$& $22.8 \pm 13$& $115 \pm 68$& $6800 \pm 400$& $3300 \pm 100$\\
\hline
Si/Si$_3$N$_4$& $0.823 \pm 0.16$& $2.02 \pm 0.24$& $6.12 \pm 6.6$& $23.2 \pm 24$& $7400 \pm 1300$& $3000 \pm 500$\\
\hline
Si& $0.194 \pm 0.029$& $1.07 \pm 0.13$& $7.92 \pm 2.2$& $15.7 \pm 4$& $27500 \pm 3000$& $10000 \pm 900$\\
\hline
\end{tabular}
\caption{Extracted parameters of the TLS bath for the three types of substrates. The confidence intervals are the standard deviation of the measurement clusters.}
\label{tab:results}
\end{center}
\end{table*}

Although recent experiments conducted on state-of-the art high-Q resonators have seen evidence of TLS-TLS interactions in the spectrum of resonator frequency fluctuations \cite{Burnett2014, Faoro2015}, or on the power-dependence of the damping rate in single-tone experiments \cite{Burnett2016}, we haven't observed such signatures in our experiments. In particular, the generalized tunneling model proposed in \cite{Faoro2012} predicts a logarithmic dependence of the resonator damping as a function of $\bar n$. The absence of such signature in the samples fabricated on amorphous substrates such as Si/SiO$_2$ or Si/Si$_3$N$_4$ is not surprising as the density of TLSs in bulk amorphous material is too low to induce strong coupling between neighboring TLSs \cite{Faoro2012}. On the other hand, the dominating TLSs for the Si substrates are likely located at the oxide interface layers and should thus induce the same non-trivial power-dependence of the damping rate. The discrepancy between our experiment and these recent works is probably due to the lack of chemical surface treatment prior to metal deposition in our sample processing. However, extending our simple model to the case of interacting TLSs could provide useful insight on the physics of these complex systems. In particular, the two-tone experiment demonstrated here may be useful to test some of the underlying hypotheses of the generalized tunneling model, such as a temperature dependent dephasing rate of the TLSs \cite{Faoro2012}.

\section{Conclusion}

In conclusion, we have presented an experimental method to characterize the non-linear properties of a TLS bath---the dominant loss channel of planar superconducting resonators. The method has been applied to lumped-element resonators that have been specifically optimized to confine the electric field in a small region around the substrate surface. By selectively saturating a fraction of the TLSs that are resonant with a detuned-pump field, and simultaneously measuring the cavity spectrum with a weak probe field, we have observed clear signatures of ``spectral hole burning'' in the TLS frequency distribution. The details of the evolution of the resonance frequencies and damping as a function of the pump detuning can be used to infer physical properties of the bath, such as its average dephasing rate $\Gamma_2$. This technique requires only standard microwave equipment and it is readily applicable to a large range of microwave resonators. \sam{Beyond their interest for the investigation of amorphous materials, the experimental signatures reported in this work can be used as a method to unambiguously identify the precise contribution of the TLS-bath to the total resonator loss, and thus gives a useful insight into the design and optimization of superconducting cavities.}

Furthermore, the physical situation considered here is ubiquitous in parametrically coupled systems where a particular interaction is activated by a strong pump field detuned with respect to the resonant mode frequency. \sam{This is notably the case in reservoir engineering, where the mode of interest inherits a non-trivial dissipation mechanism from the parametric coupling to an intentionally low-Q microwave mode \cite{Kapit2017}.} In this regime, the coupled dynamics of the resonant mode and TLS bath have to be carefully studied since pump photons can be scattered to the resonator via the interaction with TLSs, leading to an effective heating process. This phenomenon is currently under theoretical investigation \cite{Forni2018} and its characterization, which requires a quantum limited read-out to resolve the associated fluctuations, will be the subject of future work.

\section*{APPENDIX A: MAXWELL-BLOCH EQUATIONS IN THE PRESENCE OF DEPHASING}

The interaction of a single TLS with the cavity is described by the Jaynes-Cummings Hamiltonian (in a frame rotating at the pump frequency $\omega_p$):
\begin{align*}
H &= \hbar (\omega_\mathrm{p}-\omega_\mathrm{c}) \ad \an 
+ \frac{\hbar (\omega_\mathrm{p} - \omega_\mathrm{q})}{2} \sz \\
&+ i \hbar g (\ad \sn - \an \sd)
+ i\hbar J\left(\ad-\an\right),
\end{align*}
where $J = a_\mathrm{in} \Gamma_\mathrm{ext}/\sqrt{2}$  is supposed to be real without loss of generality (the factor $\sqrt{2}$ accounts for a ``hanger-type'' waveguide coupling). The dissipation is described by the Lindblad equation:
\begin{align*}
\label{Lindbladt}
\frac{d \rho}{dt} &= -\frac{i}{\hbar} [\boldsymbol{H}, \rho] + \Gamma_{\uparrow \downarrow}(\n+1) \mathcal{D}_{\sn}(\rho) +  \frac{\Gamma_\phi}{2}  \mathcal{D}_{\sz}(\rho)\\
&+\Gamma_{\uparrow \downarrow}\n\mathcal{D}_{\sd}(\rho)+\Gamma_\mathrm{ext}\mathcal{D}_{\an}(\rho),
\end{align*}
with $n_{\mathrm{th}}=\frac{1}{e^{\hbar\omega/kT}-1}$ the occupation number of the TLS, $\Gamma_\mathrm{ext}$ the damping of the cavity in the absence of TLS, $\Gamma_\phi$ the eventual TLS dephasing rate, $\Gamma_{\uparrow \downarrow}$ its energy loss rate at zero temperature, and
\[
\mathcal{D}_{\boldsymbol{A}}(\rho)=\boldsymbol{A}\rho \boldsymbol{A}\da-\frac{1}{2}(\boldsymbol{A}\da \boldsymbol{A}\rho+\rho \boldsymbol{A}\da \boldsymbol{A}).
\]
Using the formulas $\langle \boldsymbol{A} \rangle=\tr( \boldsymbol{A} \rho)$ and $\frac{d}{dt}\langle \boldsymbol{A} \rangle=\tr(\boldsymbol{A} \frac{d}{dt}\rho)$, one can compute the Maxwell-Bloch equations
\begin{align}
\label{EOM1}
\frac{d \langle \an \rangle}{dt} &=  (- i \Delta-\frac{\Gamma_\mathrm{ext}}{2}) \langle  \an \rangle  +  g\langle  \sn \rangle + J \\
\label{EOM2}
\frac{d \langle \sn \rangle }{dt} &=  \left(- i(\omega_\mathrm{p} - \omega_\mathrm{q}) - \Gamma_2\right) \langle  \sn \rangle  + g \langle \an \sz \rangle \\
\label{EOM3}
\frac{d \langle \sz \rangle }{dt} &=  -2g ( \langle \ad \sn \rangle + \langle \an \sd \rangle ) - \Gamma_1 \left(\langle \sz \rangle - \langle \sz \rangle_\mathrm{th} \right),
\end{align}
where we defined $\Gamma_2=\frac{\Gamma_{\uparrow \downarrow}}{2}(1+2\n) + \Gamma_\phi$, $\Gamma_1 = \Gamma_{\uparrow \downarrow}(1+2\n)$, $\langle \sz \rangle_\mathrm{th} = - 1/(1 + 2 \n) = - \tanh(\hbar \omega/2 k_\mathrm{B} T)$ and $\Delta=\omega_\mathrm{p}-\omega_\mathrm{c}$. 

\section{APPENDIX B: FULL  DERIVATION OF THE MODEL}

To transform this system ((\ref{EOM1}), (\ref{EOM2}), and (\ref{EOM3})) into a closed set of equations, we neglect the correlations and factorize the products $ \langle \an \sz  \rangle = \langle  \an \rangle \langle \sz  \rangle$, $\langle \ad \sn  \rangle = \langle  \ad \rangle \langle \sn  \rangle$ and $\langle \an \sd  \rangle = \langle  \an \rangle \langle \sd \rangle$. Moreover, we decompose the mean values into semi-classical stationary and modulated components $\langle \an \rangle = \alpha + \delta \alpha(t) e^{-i \Delta t}$, $\langle \sn \rangle = \sigma_0 + \delta \sigma(t) e^{-i \Delta t}$, $\langle \sz \rangle = \sigma_{z0}$, and with $\delta \alpha(t)$ and $\delta \sigma(t)$ slowly varying complex functions. The equations for the stationary components then read:
\begin{align}
\label{eq:bloch1}
0&=(-i\Delta-\Gamma_\mathrm{ext}/2)\alpha+g\sigma_0+J\\
\label{eq:bloch2}
0&=(-i(\omega_p-\omega_q)-\Gamma_2)\sigma_0+g\alpha\sigma_{\mathrm{z0}}\\
\label{eq:bloch3}
0&=-2g(\alpha^*\sigma_0+\alpha\sigma_0^*)-\Gamma_1(\sigma_{\mathrm{z0}}-  \langle \sz \rangle_\mathrm{th} ),
\end{align}
From Eq. (\ref{eq:bloch2}), we get:
\begin{equation}
\sigma_0=\frac{g\alpha\sigma_{\mathrm{z0}}}{i(\omega_\mathrm{p}-\omega_\mathrm{q})+\Gamma_2}.
\end{equation}
and from Eq. (\ref{eq:bloch3}), we obtain the population imbalance resulting from the saturation of the transition by the pump field
\begin{equation}
 \sigma_{z0}  =  \langle \sz \rangle_\mathrm{th} \left(1 - \frac{\Gamma_2^2 \bar n/n_\mathrm{s}}{(\omega_\mathrm{q} - \omega_{\text{p}})^2 + \Gamma_2^2 (1 + \bar n/n_\mathrm{s})} \right),
\end{equation}
where $\bar n=|\alpha|^2$ is the mean photon number in the cavity and $n_\mathrm{s}^{-1}=4g^2/\Gamma_1\Gamma_2$ the number of photons required to saturate the TLS transition. We then solve for the modulated parts by adiabatically eliminating the TLS dynamic ($ \mathrm{d}{\delta\sigma}/\mathrm{d}t = 0$). From Eq. (\ref{EOM1}), we obtain
\begin{equation}
\label{eq:dalpha}
 \dot {\delta \alpha} = -\Gamma_\mathrm{ext} \delta \alpha/2 + g \delta \sigma,
\end{equation}
and from Eq. (\ref{EOM2})
\begin{equation}
\label{eq:dsigma}
\delta \sigma = \frac{g \sigma_{\mathrm{z0}} \delta \alpha }{- i (\omega_\mathrm{c} - \omega_\mathrm{q}) - \Gamma_2}.
\end{equation}
By substituting Eq. (\ref{eq:dsigma}) into Eq. (\ref{eq:dalpha}), we obtain
\begin{equation}
\label{eq:dalpha_new}
\dot{\delta \alpha} = \left(-\Gamma_\mathrm{ext}/2 + \frac{g^2 \sigma_{\mathrm{z0}}}{ i (\omega_\mathrm{c} - \omega_\mathrm{q}) + \Gamma_2}\right) \delta \alpha.
\end{equation}
Hence, the complex frequency pull is given by
\begin{equation}
\delta \omega = \frac{g^2 \sigma_{\mathrm{z0}}}{ ( \omega_\mathrm{q} - \omega_\mathrm{c}) +i \Gamma_2}.
\end{equation}

In turn, the population difference $\langle \sz(\omega_{\mathrm{q}}) \rangle$ of a single TLS induces a shift of the complex cavity frequency \cite{Katz2017}
\begin{equation}
\label{TLS effect}
\delta\omega=\frac{g^2\langle\sz (\omega_{\mathrm{q}}) \rangle}{\omega_{\mathrm{q}} - \omega_\mathrm{c} + i\Gamma_2}.
\end{equation}
\noindent We also compute the total frequency shift and damping by summing the individual contributions of all the TLSs. If we assume a homogeneous distribution of frequencies of the TLS, with a density $P_0$, we have the following
\begin{align*}
\delta\omega_\mathrm{c} &= \int_{-\infty}^{\infty} d\omega_\mathrm{q} \frac{P_0}{2 \pi} \frac{g^2 \sigma_{\mathrm{z0}}}{\omega_\mathrm{q}-\omega_\mathrm{c} + i\Gamma_2} \\
 &= \int \langle \sz\rangle_\mathrm{th} \left(1 - \frac{\Gamma_2^2 \bar n/n_\mathrm{s}}{(\omega_\mathrm{q} - \omega_{\text{p}})^2 + \Gamma_2^2 (1 + \bar n/n_\mathrm{s})} \right) \\
&\times \frac{P_0 g^2}{ ( \omega_\mathrm{q} - \omega_\mathrm{c}) +i \Gamma_2} \frac{d \omega_\mathrm{q}}{2\pi}.
\end{align*}

\noindent This integral can be interpreted as the convolution product
\begin{align*}
 \delta \omega_{\mathrm{c}}(\Delta) &= \langle \sz \rangle_\mathrm{th} P_0 g^2 \times \\ &\;\left[ \left(1 - \frac{\Gamma_2^2 \bar n/n_\mathrm{s}}{\omega^2 +\Gamma_{2}^2(1+\bar n/n_\mathrm{s})} \right) \right.
\left.\otimes  \frac{1}{\omega + i \Gamma_{2}} \right] (\Delta).
\end{align*}

\noindent Using standard Fourier transforms and the convolution theorem, we derive the expression

\begin{align*}
\delta \omega_{c} &=  -P_0 g^2 \langle \sz\rangle_{\mathrm{th}}/2 \times \\
&\left( i + \frac{\bar n/n_\mathrm{s}}{\sqrt{1+\bar n/n_\mathrm{s}}} \frac{1}{\Delta/\Gamma_2  + i ( 1 + \sqrt{1+\bar n/n_\mathrm{s}})} \right).
\end{align*}
The real and double-imaginary parts of this expression, given by formula \eqref{real_part_model} and \eqref{imaginary_part_model}, correspond to the frequency shift and damping induced by the TLS bath.

\section*{Appendix C: Photon number calibration}
\begin{figure}[H]
\includegraphics[width=\linewidth]{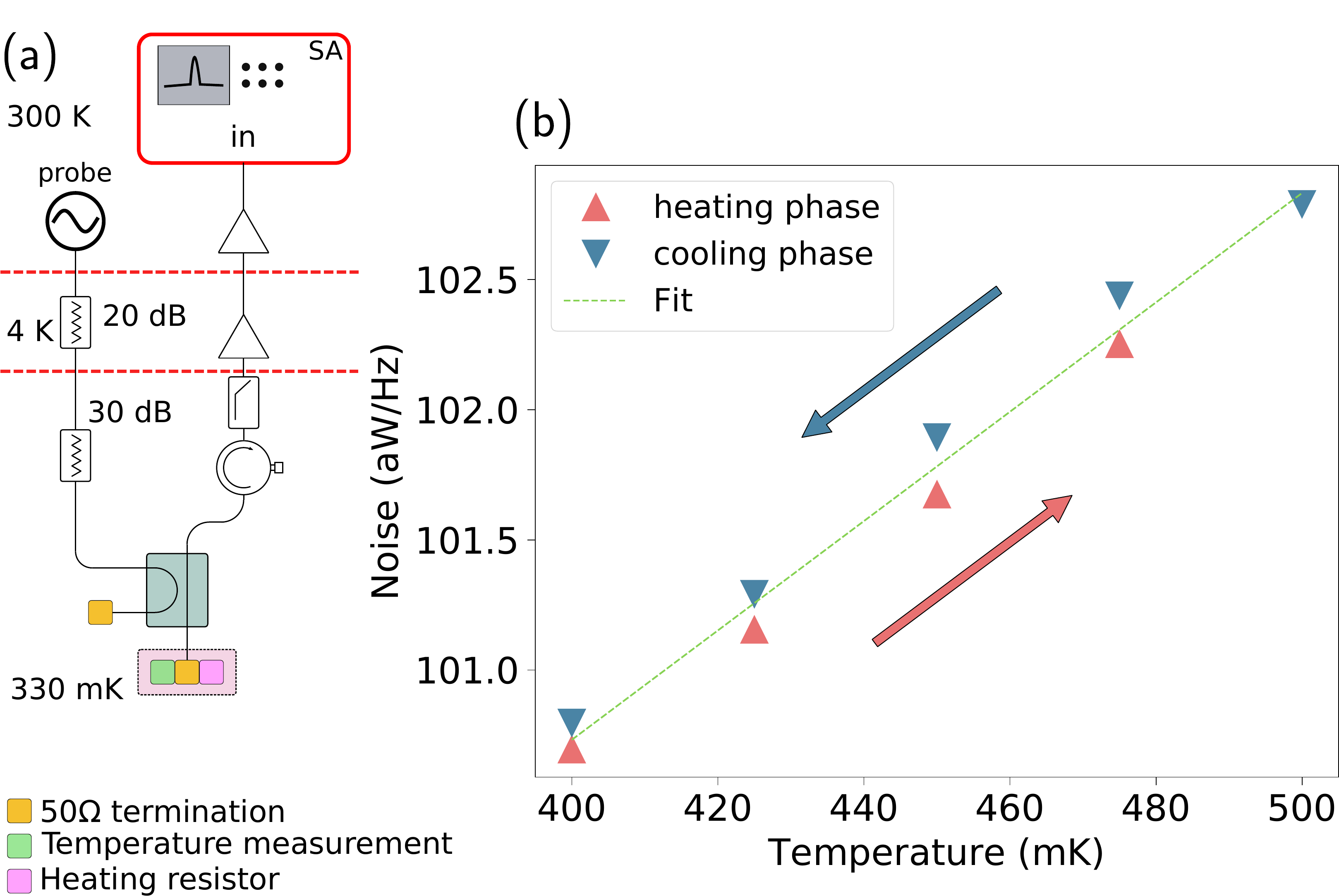}
\caption{\label{fig:thermalization}(a) Setup used for the calibration of the amplification and attenuation chain. \sam{SA, spectrum analyzer.} (b) Experimental results, for increasing and decreasing temperatures, with added fit.}
\end{figure}

The intracavity photon number at cryogenic temperature can be precisely quantified with the calibration of the total attenuation required for signal thermalization. To achieve this, we performed a separate experiment shown in Fig. \ref{fig:thermalization}(a) where a temperature controlled $50\ \Omega$-termination is placed instead of the sample. The emitted Johnson-Nyquist noise is used as a calibrated signal to  extract the gain of the amplification chain $G$. The power spectral density measured by a spectrum analyzer depends on the temperature of the $50\ \Omega$-resistor according to $S[\omega,T]=G(\hbar \omega_c[\exp(\frac{\hbar \omega_c}{k_B T}-1]^{-1}+S_\mathrm{amp})$ where $S_\mathrm{amp}$ is the spectral noise density added by the amplification chain.
The measured power spectral density is shown in Fig. \ref{fig:thermalization}(b) as a function of the temperature of the $50\ \Omega$-termination, the temperature is cycled up and down with a 5 minutes thermalization time for each value. The temperature sweep enables us to separate the contribution of the Johnson-Nyquist noise from the noise added from the amplification chain. Moreover, a probe is added to track slow drift of the gain of the amplification chain during the calibration. We obtain an amplification chain gain of $G=61.8\ \mathrm{dB}$ and an added noise corresponding to $S_\mathrm{amp}=k_B \times 4.4\ \mathrm{K}$ in agreement with the specification of the HEMT amplifier.
Assuming that the amplification chain remains unchanged with respect to the setup presented in Fig. \ref{fig:setup}(a) of the main text, we are able to extract the total attenuation of the thermalization chain of $52\ \mathrm{dB}$ and therefore infer the photon flux at the input of the superconducting resonator $|a_{\mathrm{in}}|^2$. Thus, the intracavity photon number can be extracted according to the following equation
\begin{equation}
\bar{n}=\dfrac{2\Gamma_\mathrm{ext} |a_{\mathrm{in}}|^2 }{\Gamma_{\mathrm{tot}}^2+4\Delta^2},
\end{equation}
where the coupling rate $\Gamma_\mathrm{ext}$, the total decay rate $\Gamma_\mathrm{tot}$ and the cavity detuning $\Delta$ are extracted from experimental fits such as those presented in Fig. \ref{fig:setup}(c) of the main text.

\acknowledgements{The authors would like to thank Paolo Forni and Alain Sarlette for insightful discussions, Takis Kontos for his help with the Niobium deposition facility, Xu (Sundae) Chen, Leo Morel and Leonhard Neuhaus for early developments. This work was supported by a starting grant from ''Agence Nationale pour la Recherche" QuNaT (ANR-14-CE26-0002). E.F. is funded by a Junior research chair from the LabEX ENS-ICFP: ANR-10-LABX-0010/ ANR-10-IDEX-0001-02 PSL*. E.I. acknowledges support from the European Union’s Horizon 2020 Programme for Research and Innovation under grant agreement No. 722923 (Marie Curie ETN - OMT). S. C. is supported by the grant ANR ExSqueez (ANR-15-CE30-0014).}


\begin{thebibliography}{43}%
\makeatletter
\providecommand \@ifxundefined [1]{%
 \@ifx{#1\undefined}
}%
\providecommand \@ifnum [1]{%
 \ifnum #1\expandafter \@firstoftwo
 \else \expandafter \@secondoftwo
 \fi
}%
\providecommand \@ifx [1]{%
 \ifx #1\expandafter \@firstoftwo
 \else \expandafter \@secondoftwo
 \fi
}%
\providecommand \natexlab [1]{#1}%
\providecommand \enquote  [1]{``#1''}%
\providecommand \bibnamefont  [1]{#1}%
\providecommand \bibfnamefont [1]{#1}%
\providecommand \citenamefont [1]{#1}%
\providecommand \href@noop [0]{\@secondoftwo}%
\providecommand \href [0]{\begingroup \@sanitize@url \@href}%
\providecommand \@href[1]{\@@startlink{#1}\@@href}%
\providecommand \@@href[1]{\endgroup#1\@@endlink}%
\providecommand \@sanitize@url [0]{\catcode `\\12\catcode `\$12\catcode
  `\&12\catcode `\#12\catcode `\^12\catcode `\_12\catcode `\%12\relax}%
\providecommand \@@startlink[1]{}%
\providecommand \@@endlink[0]{}%
\providecommand \url  [0]{\begingroup\@sanitize@url \@url }%
\providecommand \@url [1]{\endgroup\@href {#1}{\urlprefix }}%
\providecommand \urlprefix  [0]{URL }%
\providecommand \Eprint [0]{\href }%
\providecommand \doibase [0]{http://dx.doi.org/}%
\providecommand \selectlanguage [0]{\@gobble}%
\providecommand \bibinfo  [0]{\@secondoftwo}%
\providecommand \bibfield  [0]{\@secondoftwo}%
\providecommand \translation [1]{[#1]}%
\providecommand \BibitemOpen [0]{}%
\providecommand \bibitemStop [0]{}%
\providecommand \bibitemNoStop [0]{.\EOS\space}%
\providecommand \EOS [0]{\spacefactor3000\relax}%
\providecommand \BibitemShut  [1]{\csname bibitem#1\endcsname}%
\let\auto@bib@innerbib\@empty
\bibitem [{\citenamefont {Sauvageau}\ and\ \citenamefont
  {Macdonald}(1989)}]{Sauvageau1989}%
  \BibitemOpen
  \bibfield  {author} {\bibinfo {author} {\bibfnamefont {J.}~\bibnamefont
  {Sauvageau}}\ and\ \bibinfo {author} {\bibfnamefont {D.}~\bibnamefont
  {Macdonald}},\ }\href@noop {} {\bibfield  {journal} {\bibinfo  {journal}
  {IEEE Transactions on magnetics}\ }\textbf {\bibinfo {volume} {25}},\
  \bibinfo {pages} {1331} (\bibinfo {year} {1989})}\BibitemShut {NoStop}%
\bibitem [{\citenamefont {Audley}, \citenamefont {Kelley},\ and\ \citenamefont
  {Rawley}(1993)}]{Audley1993}%
  \BibitemOpen
  \bibfield  {author} {\bibinfo {author} {\bibfnamefont {M.~D.}\ \bibnamefont
  {Audley}}, \bibinfo {author} {\bibfnamefont {R.~L.}\ \bibnamefont {Kelley}},
  \ and\ \bibinfo {author} {\bibfnamefont {G.~L.}\ \bibnamefont {Rawley}},\
  }\href@noop {} {\bibfield  {journal} {\bibinfo  {journal} {Journal of Low
  temperature physics}\ }\textbf {\bibinfo {volume} {93}},\ \bibinfo {pages}
  {245} (\bibinfo {year} {1993})}\BibitemShut {NoStop}%
\bibitem [{\citenamefont {Roy}\ and\ \citenamefont {Devoret}(2016)}]{Roy2016}%
  \BibitemOpen
  \bibfield  {author} {\bibinfo {author} {\bibfnamefont {A.}~\bibnamefont
  {Roy}}\ and\ \bibinfo {author} {\bibfnamefont {M.}~\bibnamefont {Devoret}},\
  }\href@noop {} {\bibfield  {journal} {\bibinfo  {journal} {Comptes Rendus
  Physique}\ }\textbf {\bibinfo {volume} {17}},\ \bibinfo {pages} {740 }
  (\bibinfo {year} {2016})}\BibitemShut {NoStop}%
\bibitem [{\citenamefont {Pillet}\ \emph {et~al.}(2015)\citenamefont {Pillet},
  \citenamefont {Flurin}, \citenamefont {Mallet},\ and\ \citenamefont
  {Huard}}]{Pillet2015}%
  \BibitemOpen
  \bibfield  {author} {\bibinfo {author} {\bibfnamefont {J.-D.}\ \bibnamefont
  {Pillet}}, \bibinfo {author} {\bibfnamefont {E.}~\bibnamefont {Flurin}},
  \bibinfo {author} {\bibfnamefont {F.}~\bibnamefont {Mallet}}, \ and\ \bibinfo
  {author} {\bibfnamefont {B.}~\bibnamefont {Huard}},\ }\href@noop {}
  {\bibfield  {journal} {\bibinfo  {journal} {Applied Physics Letters}\
  }\textbf {\bibinfo {volume} {106}},\ \bibinfo {pages} {222603} (\bibinfo
  {year} {2015})}\BibitemShut {NoStop}%
\bibitem [{\citenamefont {Blais}\ \emph {et~al.}(2004)\citenamefont {Blais},
  \citenamefont {Huang}, \citenamefont {Wallraff}, \citenamefont {Girvin},\
  and\ \citenamefont {Schoelkopf}}]{Blais2004}%
  \BibitemOpen
  \bibfield  {author} {\bibinfo {author} {\bibfnamefont {A.}~\bibnamefont
  {Blais}}, \bibinfo {author} {\bibfnamefont {R.-S.}\ \bibnamefont {Huang}},
  \bibinfo {author} {\bibfnamefont {A.}~\bibnamefont {Wallraff}}, \bibinfo
  {author} {\bibfnamefont {S.~M.}\ \bibnamefont {Girvin}}, \ and\ \bibinfo
  {author} {\bibfnamefont {R.~J.}\ \bibnamefont {Schoelkopf}},\ }\href@noop {}
  {\bibfield  {journal} {\bibinfo  {journal} {Physical Review A}\ }\textbf
  {\bibinfo {volume} {69}},\ \bibinfo {pages} {062320} (\bibinfo {year}
  {2004})}\BibitemShut {NoStop}%
\bibitem [{\citenamefont {Rau}, \citenamefont {Johansson},\ and\ \citenamefont
  {Shnirman}(2004)}]{Rau2004}%
  \BibitemOpen
  \bibfield  {author} {\bibinfo {author} {\bibfnamefont {I.}~\bibnamefont
  {Rau}}, \bibinfo {author} {\bibfnamefont {G.}~\bibnamefont {Johansson}}, \
  and\ \bibinfo {author} {\bibfnamefont {A.}~\bibnamefont {Shnirman}},\
  }\href@noop {} {\bibfield  {journal} {\bibinfo  {journal} {Physical Review
  B}\ }\textbf {\bibinfo {volume} {70}},\ \bibinfo {pages} {054521} (\bibinfo
  {year} {2004})}\BibitemShut {NoStop}%
\bibitem [{\citenamefont {Wallraff}\ \emph {et~al.}(2004)\citenamefont
  {Wallraff}, \citenamefont {Schuster}, \citenamefont {Blais}, \citenamefont
  {Frunzio}, \citenamefont {Huang}, \citenamefont {Majer}, \citenamefont
  {Kumar}, \citenamefont {Girvin},\ and\ \citenamefont
  {Schoelkopf}}]{Wallraff2004}%
  \BibitemOpen
  \bibfield  {author} {\bibinfo {author} {\bibfnamefont {A.}~\bibnamefont
  {Wallraff}}, \bibinfo {author} {\bibfnamefont {D.}~\bibnamefont {Schuster}},
  \bibinfo {author} {\bibfnamefont {A.}~\bibnamefont {Blais}}, \bibinfo
  {author} {\bibfnamefont {L.}~\bibnamefont {Frunzio}}, \bibinfo {author}
  {\bibfnamefont {R.-S.}\ \bibnamefont {Huang}}, \bibinfo {author}
  {\bibfnamefont {J.}~\bibnamefont {Majer}}, \bibinfo {author} {\bibfnamefont
  {S.}~\bibnamefont {Kumar}}, \bibinfo {author} {\bibfnamefont
  {S.}~\bibnamefont {Girvin}}, \ and\ \bibinfo {author} {\bibfnamefont
  {R.}~\bibnamefont {Schoelkopf}},\ }\href@noop {} {\bibfield  {journal}
  {\bibinfo  {journal} {Nature}\ }\textbf {\bibinfo {volume} {461}},\ \bibinfo
  {pages} {162} (\bibinfo {year} {2004})}\BibitemShut {NoStop}%
\bibitem [{\citenamefont {Andr\'e}\ \emph {et~al.}(2006)\citenamefont
  {Andr\'e}, \citenamefont {DeMille}, \citenamefont {Doyle}, \citenamefont
  {Lukin}, \citenamefont {Maxwell}, \citenamefont {Rabl}, \citenamefont
  {Schoelkopf},\ and\ \citenamefont {Zoller}}]{Andre2006}%
  \BibitemOpen
  \bibfield  {author} {\bibinfo {author} {\bibfnamefont {A.}~\bibnamefont
  {Andr\'e}}, \bibinfo {author} {\bibfnamefont {D.}~\bibnamefont {DeMille}},
  \bibinfo {author} {\bibfnamefont {J.~M.}\ \bibnamefont {Doyle}}, \bibinfo
  {author} {\bibfnamefont {M.~D.}\ \bibnamefont {Lukin}}, \bibinfo {author}
  {\bibfnamefont {S.~E.}\ \bibnamefont {Maxwell}}, \bibinfo {author}
  {\bibfnamefont {P.}~\bibnamefont {Rabl}}, \bibinfo {author} {\bibfnamefont
  {R.~J.}\ \bibnamefont {Schoelkopf}}, \ and\ \bibinfo {author} {\bibfnamefont
  {P.}~\bibnamefont {Zoller}},\ }\href@noop {} {\bibfield  {journal} {\bibinfo
  {journal} {Nature Physics}\ }\textbf {\bibinfo {volume} {2}},\ \bibinfo
  {pages} {636} (\bibinfo {year} {2006})}\BibitemShut {NoStop}%
\bibitem [{\citenamefont {Teufel}\ \emph {et~al.}(2011)\citenamefont {Teufel},
  \citenamefont {Donner}, \citenamefont {Dale}, \citenamefont {Harlow},
  \citenamefont {Allman}, \citenamefont {Cicak}, \citenamefont {Sirois},
  \citenamefont {Whittaker}, \citenamefont {Lehnert},\ and\ \citenamefont
  {Simmonds}}]{Teufel2011}%
  \BibitemOpen
  \bibfield  {author} {\bibinfo {author} {\bibfnamefont {J.}~\bibnamefont
  {Teufel}}, \bibinfo {author} {\bibfnamefont {T.}~\bibnamefont {Donner}},
  \bibinfo {author} {\bibfnamefont {L.}~\bibnamefont {Dale}}, \bibinfo {author}
  {\bibfnamefont {J.}~\bibnamefont {Harlow}}, \bibinfo {author} {\bibfnamefont
  {M.~S.}\ \bibnamefont {Allman}}, \bibinfo {author} {\bibfnamefont
  {K.}~\bibnamefont {Cicak}}, \bibinfo {author} {\bibfnamefont {A.~J.}\
  \bibnamefont {Sirois}}, \bibinfo {author} {\bibfnamefont {J.~D.}\
  \bibnamefont {Whittaker}}, \bibinfo {author} {\bibfnamefont {K.~W.}\
  \bibnamefont {Lehnert}}, \ and\ \bibinfo {author} {\bibfnamefont {R.~W.}\
  \bibnamefont {Simmonds}},\ }\href@noop {} {\bibfield  {journal} {\bibinfo
  {journal} {Nature}\ }\textbf {\bibinfo {volume} {475}},\ \bibinfo {pages}
  {359} (\bibinfo {year} {2011})}\BibitemShut {NoStop}%
\bibitem [{\citenamefont {Toth}\ \emph {et~al.}(2017)\citenamefont {Toth},
  \citenamefont {Bernier}, \citenamefont {Nunnenkamp}, \citenamefont
  {Feofanov},\ and\ \citenamefont {Kippenberg}}]{Toth2017}%
  \BibitemOpen
  \bibfield  {author} {\bibinfo {author} {\bibfnamefont {L.}~\bibnamefont
  {Toth}}, \bibinfo {author} {\bibfnamefont {N.~R.}\ \bibnamefont {Bernier}},
  \bibinfo {author} {\bibfnamefont {A.}~\bibnamefont {Nunnenkamp}}, \bibinfo
  {author} {\bibfnamefont {A.~K.}\ \bibnamefont {Feofanov}}, \ and\ \bibinfo
  {author} {\bibfnamefont {T.~J.}\ \bibnamefont {Kippenberg}},\ }\href@noop {}
  {\bibfield  {journal} {\bibinfo  {journal} {Nature}\ }\textbf {\bibinfo
  {volume} {13}},\ \bibinfo {pages} {783} (\bibinfo {year} {2017})}\BibitemShut
  {NoStop}%
\bibitem [{\citenamefont {Tabuchi}\ \emph {et~al.}(2014)\citenamefont
  {Tabuchi}, \citenamefont {Ishino}, \citenamefont {Ishikawa}, \citenamefont
  {Yamazaki}, \citenamefont {Usami},\ and\ \citenamefont
  {Nakamura}}]{Tabuchi2014}%
  \BibitemOpen
  \bibfield  {author} {\bibinfo {author} {\bibfnamefont {Y.}~\bibnamefont
  {Tabuchi}}, \bibinfo {author} {\bibfnamefont {S.}~\bibnamefont {Ishino}},
  \bibinfo {author} {\bibfnamefont {T.}~\bibnamefont {Ishikawa}}, \bibinfo
  {author} {\bibfnamefont {R.}~\bibnamefont {Yamazaki}}, \bibinfo {author}
  {\bibfnamefont {K.}~\bibnamefont {Usami}}, \ and\ \bibinfo {author}
  {\bibfnamefont {Y.}~\bibnamefont {Nakamura}},\ }\href {\doibase
  10.1103/PhysRevLett.113.083603} {\bibfield  {journal} {\bibinfo  {journal}
  {Physical Review Letter}\ }\textbf {\bibinfo {volume} {113}},\ \bibinfo
  {pages} {083603} (\bibinfo {year} {2014})}\BibitemShut {NoStop}%
\bibitem [{\citenamefont {Tabuchi}\ \emph {et~al.}(2015)\citenamefont
  {Tabuchi}, \citenamefont {Ishino}, \citenamefont {Noguchi}, \citenamefont
  {Ishikawa}, \citenamefont {Yamazaki}, \citenamefont {Usami},\ and\
  \citenamefont {Nakamura}}]{Tabuchi2015}%
  \BibitemOpen
  \bibfield  {author} {\bibinfo {author} {\bibfnamefont {Y.}~\bibnamefont
  {Tabuchi}}, \bibinfo {author} {\bibfnamefont {S.}~\bibnamefont {Ishino}},
  \bibinfo {author} {\bibfnamefont {A.}~\bibnamefont {Noguchi}}, \bibinfo
  {author} {\bibfnamefont {T.}~\bibnamefont {Ishikawa}}, \bibinfo {author}
  {\bibfnamefont {R.}~\bibnamefont {Yamazaki}}, \bibinfo {author}
  {\bibfnamefont {K.}~\bibnamefont {Usami}}, \ and\ \bibinfo {author}
  {\bibfnamefont {Y.}~\bibnamefont {Nakamura}},\ }\href {\doibase
  10.1126/science.aaa3693} {\bibfield  {journal} {\bibinfo  {journal}
  {Science}\ }\textbf {\bibinfo {volume} {349}},\ \bibinfo {pages} {405}
  (\bibinfo {year} {2015})}\BibitemShut {NoStop}%
\bibitem [{\citenamefont {Wang}\ \emph {et~al.}(2009)\citenamefont {Wang},
  \citenamefont {Hofheinz}, \citenamefont {Wenner}, \citenamefont {Ansmann},
  \citenamefont {Bialczak}, \citenamefont {Lenander}, \citenamefont {Lucero},
  \citenamefont {Neeley}, \citenamefont {O’Connell}, \citenamefont {Sank},
  \citenamefont {Weides}, \citenamefont {Cleland},\ and\ \citenamefont
  {Martinis}}]{Wang2009}%
  \BibitemOpen
  \bibfield  {author} {\bibinfo {author} {\bibfnamefont {H.}~\bibnamefont
  {Wang}}, \bibinfo {author} {\bibfnamefont {M.}~\bibnamefont {Hofheinz}},
  \bibinfo {author} {\bibfnamefont {J.}~\bibnamefont {Wenner}}, \bibinfo
  {author} {\bibfnamefont {M.}~\bibnamefont {Ansmann}}, \bibinfo {author}
  {\bibfnamefont {R.~C.}\ \bibnamefont {Bialczak}}, \bibinfo {author}
  {\bibfnamefont {M.}~\bibnamefont {Lenander}}, \bibinfo {author}
  {\bibfnamefont {E.}~\bibnamefont {Lucero}}, \bibinfo {author} {\bibfnamefont
  {M.}~\bibnamefont {Neeley}}, \bibinfo {author} {\bibfnamefont {A.~D.}\
  \bibnamefont {O’Connell}}, \bibinfo {author} {\bibfnamefont
  {D.}~\bibnamefont {Sank}}, \bibinfo {author} {\bibfnamefont {M.}~\bibnamefont
  {Weides}}, \bibinfo {author} {\bibfnamefont {A.~N.}\ \bibnamefont {Cleland}},
  \ and\ \bibinfo {author} {\bibfnamefont {J.~M.}\ \bibnamefont {Martinis}},\
  }\href {\doibase 10.1063/1.3273372} {\bibfield  {journal} {\bibinfo
  {journal} {Applied Physics Letters}\ }\textbf {\bibinfo {volume} {95}},\
  \bibinfo {pages} {233508} (\bibinfo {year} {2009})}\BibitemShut {NoStop}%
\bibitem [{\citenamefont {Gao}\ \emph {et~al.}(2008)\citenamefont {Gao},
  \citenamefont {Daal}, \citenamefont {Martinis}, \citenamefont {Vayonakis},
  \citenamefont {Zmuidzinas}, \citenamefont {Sadoulet}, \citenamefont {Mazin},
  \citenamefont {Day},\ and\ \citenamefont {Leduc}}]{Gao2008}%
  \BibitemOpen
  \bibfield  {author} {\bibinfo {author} {\bibfnamefont {J.}~\bibnamefont
  {Gao}}, \bibinfo {author} {\bibfnamefont {M.}~\bibnamefont {Daal}}, \bibinfo
  {author} {\bibfnamefont {J.~M.}\ \bibnamefont {Martinis}}, \bibinfo {author}
  {\bibfnamefont {A.}~\bibnamefont {Vayonakis}}, \bibinfo {author}
  {\bibfnamefont {J.}~\bibnamefont {Zmuidzinas}}, \bibinfo {author}
  {\bibfnamefont {B.}~\bibnamefont {Sadoulet}}, \bibinfo {author}
  {\bibfnamefont {B.~A.}\ \bibnamefont {Mazin}}, \bibinfo {author}
  {\bibfnamefont {P.~K.}\ \bibnamefont {Day}}, \ and\ \bibinfo {author}
  {\bibfnamefont {H.~G.}\ \bibnamefont {Leduc}},\ }\href {\doibase
  10.1063/1.2937855} {\bibfield  {journal} {\bibinfo  {journal} {Applied
  Physics Letters}\ }\textbf {\bibinfo {volume} {92}},\ \bibinfo {pages}
  {212504} (\bibinfo {year} {2008})}\BibitemShut {NoStop}%
\bibitem [{\citenamefont {Lindstr\"{o}m}\ \emph {et~al.}(2009)\citenamefont
  {Lindstr\"{o}m}, \citenamefont {Healey}, \citenamefont {Colclough},
  \citenamefont {Muirhead},\ and\ \citenamefont {Tzalenchuk}}]{Lindstrom2009}%
  \BibitemOpen
  \bibfield  {author} {\bibinfo {author} {\bibfnamefont {T.}~\bibnamefont
  {Lindstr\"{o}m}}, \bibinfo {author} {\bibfnamefont {J.~E.}\ \bibnamefont
  {Healey}}, \bibinfo {author} {\bibfnamefont {M.~S.}\ \bibnamefont
  {Colclough}}, \bibinfo {author} {\bibfnamefont {C.~M.}\ \bibnamefont
  {Muirhead}}, \ and\ \bibinfo {author} {\bibfnamefont {A.~Y.}\ \bibnamefont
  {Tzalenchuk}},\ }\href {\doibase 10.1103/PhysRevB.80.132501} {\bibfield
  {journal} {\bibinfo  {journal} {Physical Review B}\ }\textbf {\bibinfo
  {volume} {80}},\ \bibinfo {pages} {132501} (\bibinfo {year}
  {2009})}\BibitemShut {NoStop}%
\bibitem [{\citenamefont {Skacel}\ \emph {et~al.}(2015)\citenamefont {Skacel},
  \citenamefont {Kaiser}, \citenamefont {Wuensch}, \citenamefont {Rotzinger},
  \citenamefont {Lukashenko}, \citenamefont {Jerger}, \citenamefont {Weiss},
  \citenamefont {Siegel},\ and\ \citenamefont {Ustinov}}]{Skacel2015}%
  \BibitemOpen
  \bibfield  {author} {\bibinfo {author} {\bibfnamefont {S.~T.}\ \bibnamefont
  {Skacel}}, \bibinfo {author} {\bibfnamefont {C.}~\bibnamefont {Kaiser}},
  \bibinfo {author} {\bibfnamefont {S.}~\bibnamefont {Wuensch}}, \bibinfo
  {author} {\bibfnamefont {H.}~\bibnamefont {Rotzinger}}, \bibinfo {author}
  {\bibfnamefont {A.}~\bibnamefont {Lukashenko}}, \bibinfo {author}
  {\bibfnamefont {M.}~\bibnamefont {Jerger}}, \bibinfo {author} {\bibfnamefont
  {G.}~\bibnamefont {Weiss}}, \bibinfo {author} {\bibfnamefont
  {M.}~\bibnamefont {Siegel}}, \ and\ \bibinfo {author} {\bibfnamefont {A.~V.}\
  \bibnamefont {Ustinov}},\ }\href@noop {} {\bibfield  {journal} {\bibinfo
  {journal} {Applied Physics Letters}\ }\textbf {\bibinfo {volume} {106}},\
  \bibinfo {pages} {022603} (\bibinfo {year} {2015})}\BibitemShut {NoStop}%
\bibitem [{\citenamefont {Sage}\ \emph {et~al.}(2011)\citenamefont {Sage},
  \citenamefont {Bolkhovsky}, \citenamefont {Oliver}, \citenamefont {Turek},\
  and\ \citenamefont {Welander}}]{Sage2011}%
  \BibitemOpen
  \bibfield  {author} {\bibinfo {author} {\bibfnamefont {J.~M.}\ \bibnamefont
  {Sage}}, \bibinfo {author} {\bibfnamefont {V.}~\bibnamefont {Bolkhovsky}},
  \bibinfo {author} {\bibfnamefont {W.~D.}\ \bibnamefont {Oliver}}, \bibinfo
  {author} {\bibfnamefont {B.}~\bibnamefont {Turek}}, \ and\ \bibinfo {author}
  {\bibfnamefont {P.~B.}\ \bibnamefont {Welander}},\ }\href@noop {} {\bibfield
  {journal} {\bibinfo  {journal} {Journal of Applied Physics}\ }\textbf
  {\bibinfo {volume} {109}},\ \bibinfo {pages} {063915} (\bibinfo {year}
  {2011})}\BibitemShut {NoStop}%
\bibitem [{\citenamefont {Kirsh}\ \emph {et~al.}(2017)\citenamefont {Kirsh},
  \citenamefont {Svetitsky}, \citenamefont {Burin}, \citenamefont {Schechter},\
  and\ \citenamefont {Katz}}]{Katz2017}%
  \BibitemOpen
  \bibfield  {author} {\bibinfo {author} {\bibfnamefont {N.}~\bibnamefont
  {Kirsh}}, \bibinfo {author} {\bibfnamefont {E.}~\bibnamefont {Svetitsky}},
  \bibinfo {author} {\bibfnamefont {A.~L.}\ \bibnamefont {Burin}}, \bibinfo
  {author} {\bibfnamefont {M.}~\bibnamefont {Schechter}}, \ and\ \bibinfo
  {author} {\bibfnamefont {N.}~\bibnamefont {Katz}},\ }\href {\doibase
  10.1103/PhysRevMaterials.1.012601} {\bibfield  {journal} {\bibinfo  {journal}
  {Phys. Rev. Materials}\ }\textbf {\bibinfo {volume} {1}},\ \bibinfo {pages}
  {012601} (\bibinfo {year} {2017})}\BibitemShut {NoStop}%
\bibitem [{\citenamefont {Lisenfeld}\ \emph {et~al.}(2015)\citenamefont
  {Lisenfeld}, \citenamefont {Grabovskij}, \citenamefont {M{\"u}ller},
  \citenamefont {Cole}, \citenamefont {Weiss},\ and\ \citenamefont
  {Ustinov}}]{Lisenfeld2015}%
  \BibitemOpen
  \bibfield  {author} {\bibinfo {author} {\bibfnamefont {J.}~\bibnamefont
  {Lisenfeld}}, \bibinfo {author} {\bibfnamefont {G.~J.}\ \bibnamefont
  {Grabovskij}}, \bibinfo {author} {\bibfnamefont {C.}~\bibnamefont
  {M{\"u}ller}}, \bibinfo {author} {\bibfnamefont {J.~H.}\ \bibnamefont
  {Cole}}, \bibinfo {author} {\bibfnamefont {G.}~\bibnamefont {Weiss}}, \ and\
  \bibinfo {author} {\bibfnamefont {A.~V.}\ \bibnamefont {Ustinov}},\
  }\href@noop {} {\bibfield  {journal} {\bibinfo  {journal} {Nature
  communications}\ }\textbf {\bibinfo {volume} {6}},\ \bibinfo {pages} {6182}
  (\bibinfo {year} {2015})}\BibitemShut {NoStop}%
\bibitem [{\citenamefont {Sarabi}\ \emph {et~al.}(2016)\citenamefont {Sarabi},
  \citenamefont {Ramanayaka}, \citenamefont {Burin}, \citenamefont
  {Wellstood},\ and\ \citenamefont {Osborn}}]{sarabi2016}%
  \BibitemOpen
  \bibfield  {author} {\bibinfo {author} {\bibfnamefont {B.}~\bibnamefont
  {Sarabi}}, \bibinfo {author} {\bibfnamefont {A.~N.}\ \bibnamefont
  {Ramanayaka}}, \bibinfo {author} {\bibfnamefont {A.~L.}\ \bibnamefont
  {Burin}}, \bibinfo {author} {\bibfnamefont {F.~C.}\ \bibnamefont
  {Wellstood}}, \ and\ \bibinfo {author} {\bibfnamefont {K.~D.}\ \bibnamefont
  {Osborn}},\ }\href@noop {} {\bibfield  {journal} {\bibinfo  {journal}
  {Physical Review Retters}\ }\textbf {\bibinfo {volume} {116}},\ \bibinfo
  {pages} {167002} (\bibinfo {year} {2016})}\BibitemShut {NoStop}%
\bibitem [{\citenamefont {O'Connell}\ \emph {et~al.}(2008)\citenamefont
  {O'Connell}, \citenamefont {Ansmann}, \citenamefont {Bialczak}, \citenamefont
  {Hofheinz}, \citenamefont {Katz}, \citenamefont {Lucero}, \citenamefont
  {McKenney}, \citenamefont {Neeley}, \citenamefont {Wang}, \citenamefont
  {Weig}, \citenamefont {Cleland},\ and\ \citenamefont
  {Martinis}}]{OConnel2008}%
  \BibitemOpen
  \bibfield  {author} {\bibinfo {author} {\bibfnamefont {A.~D.}\ \bibnamefont
  {O'Connell}}, \bibinfo {author} {\bibfnamefont {M.}~\bibnamefont {Ansmann}},
  \bibinfo {author} {\bibfnamefont {R.~C.}\ \bibnamefont {Bialczak}}, \bibinfo
  {author} {\bibfnamefont {M.}~\bibnamefont {Hofheinz}}, \bibinfo {author}
  {\bibfnamefont {N.}~\bibnamefont {Katz}}, \bibinfo {author} {\bibfnamefont
  {E.}~\bibnamefont {Lucero}}, \bibinfo {author} {\bibfnamefont
  {C.}~\bibnamefont {McKenney}}, \bibinfo {author} {\bibfnamefont
  {M.}~\bibnamefont {Neeley}}, \bibinfo {author} {\bibfnamefont
  {H.}~\bibnamefont {Wang}}, \bibinfo {author} {\bibfnamefont {E.~M.}\
  \bibnamefont {Weig}}, \bibinfo {author} {\bibfnamefont {A.~N.}\ \bibnamefont
  {Cleland}}, \ and\ \bibinfo {author} {\bibfnamefont {J.~M.}\ \bibnamefont
  {Martinis}},\ }\href@noop {} {\bibfield  {journal} {\bibinfo  {journal}
  {Applied Physics Letters}\ }\textbf {\bibinfo {volume} {92}},\ \bibinfo
  {pages} {112903} (\bibinfo {year} {2008})}\BibitemShut {NoStop}%
\bibitem [{\citenamefont {{Khalil}}, \citenamefont {{Wellstood}},\ and\
  \citenamefont {{Osborn}}(2011)}]{Khalil2011}%
  \BibitemOpen
  \bibfield  {author} {\bibinfo {author} {\bibfnamefont {M.~S.}\ \bibnamefont
  {{Khalil}}}, \bibinfo {author} {\bibfnamefont {F.~C.}\ \bibnamefont
  {{Wellstood}}}, \ and\ \bibinfo {author} {\bibfnamefont {K.~D.}\ \bibnamefont
  {{Osborn}}},\ }\href@noop {} {\bibfield  {journal} {\bibinfo  {journal} {IEEE
  Transactions on Applied Superconductivity}\ }\textbf {\bibinfo {volume}
  {21}},\ \bibinfo {pages} {879} (\bibinfo {year} {2011})}\BibitemShut
  {NoStop}%
\bibitem [{\citenamefont {Vissers}\ \emph {et~al.}(2012)\citenamefont
  {Vissers}, \citenamefont {Kline}, \citenamefont {Gao}, \citenamefont
  {Wisbey},\ and\ \citenamefont {Pappas}}]{Vissers2012}%
  \BibitemOpen
  \bibfield  {author} {\bibinfo {author} {\bibfnamefont {M.~R.}\ \bibnamefont
  {Vissers}}, \bibinfo {author} {\bibfnamefont {J.~S.}\ \bibnamefont {Kline}},
  \bibinfo {author} {\bibfnamefont {J.}~\bibnamefont {Gao}}, \bibinfo {author}
  {\bibfnamefont {D.~S.}\ \bibnamefont {Wisbey}}, \ and\ \bibinfo {author}
  {\bibfnamefont {D.~P.}\ \bibnamefont {Pappas}},\ }\href@noop {} {\bibfield
  {journal} {\bibinfo  {journal} {Applied Physics Letters}\ }\textbf {\bibinfo
  {volume} {100}},\ \bibinfo {pages} {082602} (\bibinfo {year}
  {2012})}\BibitemShut {NoStop}%
\bibitem [{\citenamefont {Paik}\ and\ \citenamefont {Osborn}(2010)}]{Paik2010}%
  \BibitemOpen
  \bibfield  {author} {\bibinfo {author} {\bibfnamefont {H.}~\bibnamefont
  {Paik}}\ and\ \bibinfo {author} {\bibfnamefont {K.~D.}\ \bibnamefont
  {Osborn}},\ }\href@noop {} {\bibfield  {journal} {\bibinfo  {journal}
  {Applied Physics Letters}\ }\textbf {\bibinfo {volume} {96}},\ \bibinfo
  {pages} {072505} (\bibinfo {year} {2010})}\BibitemShut {NoStop}%
\bibitem [{\citenamefont {Wenner}\ \emph {et~al.}(2011)\citenamefont {Wenner},
  \citenamefont {Barends}, \citenamefont {Bialczak}, \citenamefont {Chen},
  \citenamefont {Kelly}, \citenamefont {Lucero}, \citenamefont {Mariantoni},
  \citenamefont {Megrant}, \citenamefont {OMalley}, \citenamefont {Sank},
  \citenamefont {Vainsencher}, \citenamefont {Wang}, \citenamefont {White},
  \citenamefont {Yin}, \citenamefont {Zhao}, \citenamefont {Cleland},\ and\
  \citenamefont {Martinis}}]{Wenner2011}%
  \BibitemOpen
  \bibfield  {author} {\bibinfo {author} {\bibfnamefont {J.}~\bibnamefont
  {Wenner}}, \bibinfo {author} {\bibfnamefont {R.}~\bibnamefont {Barends}},
  \bibinfo {author} {\bibfnamefont {R.~C.}\ \bibnamefont {Bialczak}}, \bibinfo
  {author} {\bibfnamefont {Y.}~\bibnamefont {Chen}}, \bibinfo {author}
  {\bibfnamefont {J.}~\bibnamefont {Kelly}}, \bibinfo {author} {\bibfnamefont
  {E.}~\bibnamefont {Lucero}}, \bibinfo {author} {\bibfnamefont
  {M.}~\bibnamefont {Mariantoni}}, \bibinfo {author} {\bibfnamefont
  {A.}~\bibnamefont {Megrant}}, \bibinfo {author} {\bibfnamefont {P.~J.~J.}\
  \bibnamefont {OMalley}}, \bibinfo {author} {\bibfnamefont
  {D.}~\bibnamefont {Sank}}, \bibinfo {author} {\bibfnamefont {A.}~\bibnamefont
  {Vainsencher}}, \bibinfo {author} {\bibfnamefont {H.}~\bibnamefont {Wang}},
  \bibinfo {author} {\bibfnamefont {T.~C.}\ \bibnamefont {White}}, \bibinfo
  {author} {\bibfnamefont {Y.}~\bibnamefont {Yin}}, \bibinfo {author}
  {\bibfnamefont {J.}~\bibnamefont {Zhao}}, \bibinfo {author} {\bibfnamefont
  {A.N.}\ \bibnamefont {Cleland}}, \ and\ \bibinfo {author} {\bibfnamefont
  {J.M.}\ \bibnamefont {Martinis}},\ }\href@noop {} {\bibfield  {journal}
  {\bibinfo  {journal} {Applied Physics Letters}\ }\textbf {\bibinfo {volume}
  {99}},\ \bibinfo {pages} {113513} (\bibinfo {year} {2011})}\BibitemShut
  {NoStop}%
\bibitem [{\citenamefont {Quintana}\ \emph {et~al.}(2014)\citenamefont
  {Quintana}, \citenamefont {Megrant}, \citenamefont {Chen}, \citenamefont
  {Dunsworth}, \citenamefont {Chiaro}, \citenamefont {Barends}, \citenamefont
  {Campbell}, \citenamefont {Chen}, \citenamefont {Hoi}, \citenamefont
  {Jeffrey}, \citenamefont {Kelly}, \citenamefont {Mutus}, \citenamefont
  {O'Malley}, \citenamefont {Neill}, \citenamefont {Roushan}, \citenamefont
  {Sank}, \citenamefont {Vainsencher}, \citenamefont {Wenner}, \citenamefont
  {White}, \citenamefont {Cleland},\ and\ \citenamefont
  {Martinis}}]{Quintana2014}%
  \BibitemOpen
  \bibfield  {author} {\bibinfo {author} {\bibfnamefont {C.~M.}\ \bibnamefont
  {Quintana}}, \bibinfo {author} {\bibfnamefont {A.}~\bibnamefont {Megrant}},
  \bibinfo {author} {\bibfnamefont {Z.}~\bibnamefont {Chen}}, \bibinfo {author}
  {\bibfnamefont {A.}~\bibnamefont {Dunsworth}}, \bibinfo {author}
  {\bibfnamefont {B.}~\bibnamefont {Chiaro}}, \bibinfo {author} {\bibfnamefont
  {R.}~\bibnamefont {Barends}}, \bibinfo {author} {\bibfnamefont
  {B.}~\bibnamefont {Campbell}}, \bibinfo {author} {\bibfnamefont
  {Y.}~\bibnamefont {Chen}}, \bibinfo {author} {\bibfnamefont {I.-C.}\
  \bibnamefont {Hoi}}, \bibinfo {author} {\bibfnamefont {E.}~\bibnamefont
  {Jeffrey}}, \bibinfo {author} {\bibfnamefont {J.}~\bibnamefont {Kelly}},
  \bibinfo {author} {\bibfnamefont {J.~Y.}\ \bibnamefont {Mutus}}, \bibinfo
  {author} {\bibfnamefont {P.~J.~J.}\ \bibnamefont {O'Malley}}, \bibinfo
  {author} {\bibfnamefont {C.}~\bibnamefont {Neill}}, \bibinfo {author}
  {\bibfnamefont {P.}~\bibnamefont {Roushan}}, \bibinfo {author} {\bibfnamefont
  {D.}~\bibnamefont {Sank}}, \bibinfo {author} {\bibfnamefont {A.}~\bibnamefont
  {Vainsencher}}, \bibinfo {author} {\bibfnamefont {J.}~\bibnamefont {Wenner}},
  \bibinfo {author} {\bibfnamefont {T.~C.}\ \bibnamefont {White}}, \bibinfo
  {author} {\bibfnamefont {A.~N.}\ \bibnamefont {Cleland}}, \ and\ \bibinfo
  {author} {\bibfnamefont {J.~M.}\ \bibnamefont {Martinis}},\ }\href@noop {}
  {\bibfield  {journal} {\bibinfo  {journal} {Applied Physics Letters}\
  }\textbf {\bibinfo {volume} {105}},\ \bibinfo {pages} {062601} (\bibinfo
  {year} {2014})}\BibitemShut {NoStop}%
\bibitem [{\citenamefont {Bruno}\ \emph {et~al.}(2015)\citenamefont {Bruno},
  \citenamefont {Lange}, \citenamefont {Asaad}, \citenamefont {Enden},
  \citenamefont {Langford},\ and\ \citenamefont {Dicarlo}}]{Bruno2015}%
  \BibitemOpen
  \bibfield  {author} {\bibinfo {author} {\bibfnamefont {A.}~\bibnamefont
  {Bruno}}, \bibinfo {author} {\bibfnamefont {G.~D.}\ \bibnamefont {Lange}},
  \bibinfo {author} {\bibfnamefont {S.}~\bibnamefont {Asaad}}, \bibinfo
  {author} {\bibfnamefont {K.~L. V.~D.}\ \bibnamefont {Enden}}, \bibinfo
  {author} {\bibfnamefont {N.~K.}\ \bibnamefont {Langford}}, \ and\ \bibinfo
  {author} {\bibfnamefont {L.}~\bibnamefont {Dicarlo}},\ }\href {\doibase
  10.1063/1.4919761} {\bibfield  {journal} {\bibinfo  {journal} {Appl. Phys.
  Lett.}\ }\textbf {\bibinfo {volume} {182601}},\ \bibinfo {pages} {4}
  (\bibinfo {year} {2015})}\BibitemShut {NoStop}%
\bibitem [{\citenamefont {Calusine}\ \emph {et~al.}(2018)\citenamefont
  {Calusine}, \citenamefont {Melville}, \citenamefont {Woods}, \citenamefont
  {Das}, \citenamefont {Stull}, \citenamefont {Bolkhovsky}, \citenamefont
  {Braje}, \citenamefont {Hover}, \citenamefont {Kim}, \citenamefont {Miloshi},
  \citenamefont {Rosenberg}, \citenamefont {Sevi}, \citenamefont {Yoder},
  \citenamefont {Dauler},\ and\ \citenamefont {Oliver}}]{Calusine2018}%
  \BibitemOpen
  \bibfield  {author} {\bibinfo {author} {\bibfnamefont {G.}~\bibnamefont
  {Calusine}}, \bibinfo {author} {\bibfnamefont {A.}~\bibnamefont {Melville}},
  \bibinfo {author} {\bibfnamefont {W.}~\bibnamefont {Woods}}, \bibinfo
  {author} {\bibfnamefont {R.}~\bibnamefont {Das}}, \bibinfo {author}
  {\bibfnamefont {C.}~\bibnamefont {Stull}}, \bibinfo {author} {\bibfnamefont
  {V.}~\bibnamefont {Bolkhovsky}}, \bibinfo {author} {\bibfnamefont
  {D.}~\bibnamefont {Braje}}, \bibinfo {author} {\bibfnamefont
  {D.}~\bibnamefont {Hover}}, \bibinfo {author} {\bibfnamefont {D.~K.}\
  \bibnamefont {Kim}}, \bibinfo {author} {\bibfnamefont {X.}~\bibnamefont
  {Miloshi}}, \bibinfo {author} {\bibfnamefont {D.}~\bibnamefont {Rosenberg}},
  \bibinfo {author} {\bibfnamefont {A.}~\bibnamefont {Sevi}}, \bibinfo {author}
  {\bibfnamefont {J.~L.}\ \bibnamefont {Yoder}}, \bibinfo {author}
  {\bibfnamefont {E.}~\bibnamefont {Dauler}}, \ and\ \bibinfo {author}
  {\bibfnamefont {W.~D.}\ \bibnamefont {Oliver}},\ }\href@noop {} {\bibfield
  {journal} {\bibinfo  {journal} {Applied Physics Letters}\ }\textbf {\bibinfo
  {volume} {112}},\ \bibinfo {pages} {062601} (\bibinfo {year}
  {2018})}\BibitemShut {NoStop}%
\bibitem [{\citenamefont {Leghtas}\ \emph {et~al.}(2015)\citenamefont
  {Leghtas}, \citenamefont {Touzard}, \citenamefont {Pop}, \citenamefont {Kou},
  \citenamefont {Vlastakis}, \citenamefont {Petrenko}, \citenamefont {Sliwa},
  \citenamefont {Narla}, \citenamefont {Shankar}, \citenamefont {Hatridge},
  \citenamefont {Reagor}, \citenamefont {Frunzio}, \citenamefont {Schoelkopf},
  \citenamefont {Mirrahimi},\ and\ \citenamefont {Devoret}}]{Leghtas2015}%
  \BibitemOpen
  \bibfield  {author} {\bibinfo {author} {\bibfnamefont {Z.}~\bibnamefont
  {Leghtas}}, \bibinfo {author} {\bibfnamefont {S.}~\bibnamefont {Touzard}},
  \bibinfo {author} {\bibfnamefont {I.~M.}\ \bibnamefont {Pop}}, \bibinfo
  {author} {\bibfnamefont {A.}~\bibnamefont {Kou}}, \bibinfo {author}
  {\bibfnamefont {B.}~\bibnamefont {Vlastakis}}, \bibinfo {author}
  {\bibfnamefont {A.}~\bibnamefont {Petrenko}}, \bibinfo {author}
  {\bibfnamefont {K.~M.}\ \bibnamefont {Sliwa}}, \bibinfo {author}
  {\bibfnamefont {A.}~\bibnamefont {Narla}}, \bibinfo {author} {\bibfnamefont
  {S.}~\bibnamefont {Shankar}}, \bibinfo {author} {\bibfnamefont {M.~J.}\
  \bibnamefont {Hatridge}}, \bibinfo {author} {\bibfnamefont {M.}~\bibnamefont
  {Reagor}}, \bibinfo {author} {\bibfnamefont {L.}~\bibnamefont {Frunzio}},
  \bibinfo {author} {\bibfnamefont {R.~J.}\ \bibnamefont {Schoelkopf}},
  \bibinfo {author} {\bibfnamefont {M.}~\bibnamefont {Mirrahimi}}, \ and\
  \bibinfo {author} {\bibfnamefont {M.~H.}\ \bibnamefont {Devoret}},\ }\href
  {\doibase 10.1126/science.aaa2085} {\bibfield  {journal} {\bibinfo  {journal}
  {Science}\ }\textbf {\bibinfo {volume} {347}},\ \bibinfo {pages} {853}
  (\bibinfo {year} {2015})}\BibitemShut {NoStop}%
\bibitem [{\citenamefont {Lescanne}\ \emph {et~al.}(2019)\citenamefont
  {Lescanne}, \citenamefont {Del{\'{e}}glise}, \citenamefont {Albertinale},
  \citenamefont {R{\'{e}}glade}, \citenamefont {Ivanov}, \citenamefont
  {Jacqmin}, \citenamefont {Leghtas},\ and\ \citenamefont
  {Flurin}}]{Lescanne2019}%
  \BibitemOpen
  \bibfield  {author} {\bibinfo {author} {\bibfnamefont {R.}~\bibnamefont
  {Lescanne}}, \bibinfo {author} {\bibfnamefont {S.}~\bibnamefont
  {Del{\'{e}}glise}}, \bibinfo {author} {\bibfnamefont {E.}~\bibnamefont
  {Albertinale}}, \bibinfo {author} {\bibfnamefont {U.}~\bibnamefont
  {R{\'{e}}glade}}, \bibinfo {author} {\bibfnamefont {E.}~\bibnamefont
  {Ivanov}}, \bibinfo {author} {\bibfnamefont {T.}~\bibnamefont {Jacqmin}},
  \bibinfo {author} {\bibfnamefont {Z.}~\bibnamefont {Leghtas}}, \ and\
  \bibinfo {author} {\bibfnamefont {E.}~\bibnamefont {Flurin}},\ }\href@noop {}
  {\  (\bibinfo {year} {2019})},\ \Eprint {http://arxiv.org/abs/1902.05102}
  {arXiv:1902.05102 [quant-ph]} \BibitemShut {NoStop}%
\bibitem [{\citenamefont {Douce}\ \emph {et~al.}(2015)\citenamefont {Douce},
  \citenamefont {Stern}, \citenamefont {Zagury}, \citenamefont {Bertet},\ and\
  \citenamefont {Milman}}]{Douce2015}%
  \BibitemOpen
  \bibfield  {author} {\bibinfo {author} {\bibfnamefont {T.}~\bibnamefont
  {Douce}}, \bibinfo {author} {\bibfnamefont {M.}~\bibnamefont {Stern}},
  \bibinfo {author} {\bibfnamefont {N.}~\bibnamefont {Zagury}}, \bibinfo
  {author} {\bibfnamefont {P.}~\bibnamefont {Bertet}}, \ and\ \bibinfo {author}
  {\bibfnamefont {P.}~\bibnamefont {Milman}},\ }\href@noop {} {\bibfield
  {journal} {\bibinfo  {journal} {Physical Review A}\ }\textbf {\bibinfo
  {volume} {92}},\ \bibinfo {pages} {052335} (\bibinfo {year}
  {2015})}\BibitemShut {NoStop}%
\bibitem [{\citenamefont {Haikka}\ \emph {et~al.}(2017)\citenamefont {Haikka},
  \citenamefont {Kubo}, \citenamefont {Bienfait}, \citenamefont {Bertet},\ and\
  \citenamefont {M\o{}lmer}}]{Bertet2017}%
  \BibitemOpen
  \bibfield  {author} {\bibinfo {author} {\bibfnamefont {P.}~\bibnamefont
  {Haikka}}, \bibinfo {author} {\bibfnamefont {Y.}~\bibnamefont {Kubo}},
  \bibinfo {author} {\bibfnamefont {A.}~\bibnamefont {Bienfait}}, \bibinfo
  {author} {\bibfnamefont {P.}~\bibnamefont {Bertet}}, \ and\ \bibinfo {author}
  {\bibfnamefont {K.}~\bibnamefont {M\o{}lmer}},\ }\href {\doibase
  10.1103/PhysRevA.95.022306} {\bibfield  {journal} {\bibinfo  {journal}
  {Physycal Review A}\ }\textbf {\bibinfo {volume} {95}},\ \bibinfo {pages}
  {022306} (\bibinfo {year} {2017})}\BibitemShut {NoStop}%
\bibitem [{\citenamefont {Unterreithmeier}, \citenamefont {Weig},\ and\
  \citenamefont {Kotthaus}(2009)}]{Weig2009}%
  \BibitemOpen
  \bibfield  {author} {\bibinfo {author} {\bibfnamefont {Q.~P.}\ \bibnamefont
  {Unterreithmeier}}, \bibinfo {author} {\bibfnamefont {E.~M.}\ \bibnamefont
  {Weig}}, \ and\ \bibinfo {author} {\bibfnamefont {J.~P.}\ \bibnamefont
  {Kotthaus}},\ }\href@noop {} {\bibfield  {journal} {\bibinfo  {journal}
  {Nature}\ }\textbf {\bibinfo {volume} {458}},\ \bibinfo {pages} {1001}
  (\bibinfo {year} {2009})}\BibitemShut {NoStop}%
\bibitem [{\citenamefont {Den~Otter}(2002)}]{DenOtter2002}%
  \BibitemOpen
  \bibfield  {author} {\bibinfo {author} {\bibfnamefont {M.}~\bibnamefont
  {Den~Otter}},\ }\href@noop {} {\bibfield  {journal} {\bibinfo  {journal}
  {Sensors and Actuators, A}\ }\textbf {\bibinfo {volume} {96}},\ \bibinfo
  {pages} {140} (\bibinfo {year} {2002})}\BibitemShut {NoStop}%
\bibitem [{\citenamefont {Geerlings}\ \emph {et~al.}(2012)\citenamefont
  {Geerlings}, \citenamefont {Shankar}, \citenamefont {Edwards}, \citenamefont
  {Frunzio}, \citenamefont {Schoelkopf},\ and\ \citenamefont
  {Devoret}}]{Geerlings2012}%
  \BibitemOpen
  \bibfield  {author} {\bibinfo {author} {\bibfnamefont {K.}~\bibnamefont
  {Geerlings}}, \bibinfo {author} {\bibfnamefont {S.}~\bibnamefont {Shankar}},
  \bibinfo {author} {\bibfnamefont {E.}~\bibnamefont {Edwards}}, \bibinfo
  {author} {\bibfnamefont {L.}~\bibnamefont {Frunzio}}, \bibinfo {author}
  {\bibfnamefont {R.~J.}\ \bibnamefont {Schoelkopf}}, \ and\ \bibinfo {author}
  {\bibfnamefont {M.~H.}\ \bibnamefont {Devoret}},\ }\href {\doibase
  10.1063/1.4710520} {\bibfield  {journal} {\bibinfo  {journal} {Applied
  Physics Letters}\ }\textbf {\bibinfo {volume} {100}},\ \bibinfo {pages}
  {192601} (\bibinfo {year} {2012})}\BibitemShut {NoStop}%
\bibitem [{\citenamefont {Khalil}\ \emph {et~al.}(2012)\citenamefont {Khalil},
  \citenamefont {Stoutimore}, \citenamefont {Wellstood},\ and\ \citenamefont
  {Osborn}}]{Khalil2012}%
  \BibitemOpen
  \bibfield  {author} {\bibinfo {author} {\bibfnamefont {M.~S.}\ \bibnamefont
  {Khalil}}, \bibinfo {author} {\bibfnamefont {M.~J.~A.}\ \bibnamefont
  {Stoutimore}}, \bibinfo {author} {\bibfnamefont {F.~C.}\ \bibnamefont
  {Wellstood}}, \ and\ \bibinfo {author} {\bibfnamefont {K.~D.}\ \bibnamefont
  {Osborn}},\ }\href@noop {} {\bibfield  {journal} {\bibinfo  {journal}
  {Journal of Applied Physics}\ }\textbf {\bibinfo {volume} {111}},\ \bibinfo
  {pages} {054510} (\bibinfo {year} {2012})}\BibitemShut {NoStop}%
\bibitem [{\citenamefont {Phillips}()}]{Phillips1987}%
  \BibitemOpen
  \bibfield  {author} {\bibinfo {author} {\bibfnamefont {W.~A.}\ \bibnamefont
  {Phillips}},\ }\href@noop {} {\bibfield  {journal} {\bibinfo  {journal}
  {Report on Progress in Physics}\ }\textbf {\bibinfo {volume} {50}},\ \bibinfo
  {pages} {1657}}\BibitemShut {NoStop}%
\bibitem [{\citenamefont {Burnett}, \citenamefont {Faoro},\ and\ \citenamefont
  {Lindstr\"{o}m}(2016)}]{Burnett2016}%
  \BibitemOpen
  \bibfield  {author} {\bibinfo {author} {\bibfnamefont {J.}~\bibnamefont
  {Burnett}}, \bibinfo {author} {\bibfnamefont {L.}~\bibnamefont {Faoro}}, \
  and\ \bibinfo {author} {\bibfnamefont {T.}~\bibnamefont {Lindstr\"{o}m}},\
  }\href {\doibase 10.1088/0953-2048/29/4/044008} {\bibfield  {journal}
  {\bibinfo  {journal} {Superconductor Science and Technology}\ }\textbf
  {\bibinfo {volume} {29}},\ \bibinfo {pages} {044008} (\bibinfo {year}
  {2016})}\BibitemShut {NoStop}%
\bibitem [{\citenamefont {Burnett}\ \emph {et~al.}(2014)\citenamefont
  {Burnett}, \citenamefont {Faoro}, \citenamefont {Wisby}, \citenamefont
  {Gurtovoi}, \citenamefont {Chernykh}, \citenamefont {Mikhailov},
  \citenamefont {Tulin}, \citenamefont {Shaikhaidarov}, \citenamefont
  {Antonov}, \citenamefont {Meeson},\ and\ \citenamefont
  {Lindstr\"om}}]{Burnett2014}%
  \BibitemOpen
  \bibfield  {author} {\bibinfo {author} {\bibfnamefont {J.}~\bibnamefont
  {Burnett}}, \bibinfo {author} {\bibfnamefont {L.}~\bibnamefont {Faoro}},
  \bibinfo {author} {\bibfnamefont {I.}~\bibnamefont {Wisby}}, \bibinfo
  {author} {\bibfnamefont {V.~L.}\ \bibnamefont {Gurtovoi}}, \bibinfo {author}
  {\bibfnamefont {A.~V.}\ \bibnamefont {Chernykh}}, \bibinfo {author}
  {\bibfnamefont {G.~M.}\ \bibnamefont {Mikhailov}}, \bibinfo {author}
  {\bibfnamefont {V.~A.}\ \bibnamefont {Tulin}}, \bibinfo {author}
  {\bibfnamefont {R.}~\bibnamefont {Shaikhaidarov}}, \bibinfo {author}
  {\bibfnamefont {V.}~\bibnamefont {Antonov}}, \bibinfo {author} {\bibfnamefont
  {P.~J.}\ \bibnamefont {Meeson}}, \ and\ \bibinfo {author} {\bibfnamefont
  {T.}~\bibnamefont {Lindstr\"om}},\ }\href {\doibase 10.1038/ncomms5119}
  {\bibfield  {journal} {\bibinfo  {journal} {Nature communications}\ }\textbf
  {\bibinfo {volume} {5}},\ \bibinfo {pages} {4119} (\bibinfo {year}
  {2014})}\BibitemShut {NoStop}%
\bibitem [{\citenamefont {Faoro}\ and\ \citenamefont
  {Ioffe}(2015)}]{Faoro2015}%
  \BibitemOpen
  \bibfield  {author} {\bibinfo {author} {\bibfnamefont {L.}~\bibnamefont
  {Faoro}}\ and\ \bibinfo {author} {\bibfnamefont {L.~B.}\ \bibnamefont
  {Ioffe}},\ }\href {\doibase 10.1103/PhysRevB.91.014201} {\bibfield  {journal}
  {\bibinfo  {journal} {Physical Review B}\ }\textbf {\bibinfo {volume}
  {91}},\ \bibinfo {pages} {014201} (\bibinfo {year} {2015})}\BibitemShut
  {NoStop}%
\bibitem [{\citenamefont {Faoro}\ and\ \citenamefont
  {Ioffe}(2012)}]{Faoro2012}%
  \BibitemOpen
  \bibfield  {author} {\bibinfo {author} {\bibfnamefont {L.}~\bibnamefont
  {Faoro}}\ and\ \bibinfo {author} {\bibfnamefont {L.~B.}\ \bibnamefont
  {Ioffe}},\ }\href {\doibase 10.1103/PhysRevLett.109.157005} {\bibfield
  {journal} {\bibinfo  {journal} {Phys. Rev. Lett.}\ }\textbf {\bibinfo
  {volume} {109}},\ \bibinfo {pages} {157005} (\bibinfo {year} {2012})}\BibitemShut
  {NoStop}%
\bibitem [{\citenamefont {Kapit}(2017)}]{Kapit2017}%
  \BibitemOpen
  \bibfield  {author} {\bibinfo {author} {\bibfnamefont {E.}~\bibnamefont
  {Kapit}},\ }\href {\doibase 10.1088/2058-9565/aa7e5d} {\bibfield  {journal}
  {\bibinfo  {journal} {Quantum Science and Technology}\ }\textbf {\bibinfo
  {volume} {2}},\ \bibinfo {pages} {033002} (\bibinfo {year}
  {2017})}\BibitemShut {NoStop}%
\bibitem [{\citenamefont {Forni}\ \emph {et~al.}(2018)\citenamefont {Forni},
  \citenamefont {Sarlette}, \citenamefont {Capelle}, \citenamefont {Flurin},
  \citenamefont {Deléglise},\ and\ \citenamefont {Rouchon}}]{Forni2018}%
  \BibitemOpen
  \bibfield  {author} {\bibinfo {author} {\bibfnamefont {P.}~\bibnamefont
  {Forni}}, \bibinfo {author} {\bibfnamefont {A.}~\bibnamefont {Sarlette}},
  \bibinfo {author} {\bibfnamefont {T.}~\bibnamefont {Capelle}}, \bibinfo
  {author} {\bibfnamefont {E.}~\bibnamefont {Flurin}}, \bibinfo {author}
  {\bibfnamefont {S.}~\bibnamefont {Deléglise}}, \ and\ \bibinfo {author}
  {\bibfnamefont {P.}~\bibnamefont {Rouchon}},\ }\href@noop {} {\  (\bibinfo
  {year} {2018})},\ \Eprint {http://arxiv.org/abs/1803.07810} {arXiv:1803.07810
  [quant-ph]} \BibitemShut {NoStop}%
\end{thebibliography}

%
\end{document}